\documentclass[a4paper, 10pt, conference]{ieeeconf}      

\IEEEoverridecommandlockouts                              

\usepackage{graphics} 
\usepackage{graphicx}
\usepackage{epsfig} 
\usepackage{mathptmx} 
\usepackage{times} 

\usepackage{amsmath} 
\usepackage{amssymb}  
\usepackage{url}

\usepackage{algorithm}
\usepackage{algorithmic}
\usepackage{booktabs}
\usepackage{multirow}

\title{\LARGE \bf
Investigating the Impact of Subgraph Social Structure Preference on the Strategic Behavior of Networked Mixed-Motive Learning Agents
}

\author{Xinqi Gao$^{1}$ and Mario Ventresca$^{2}$
\thanks{This work has been submitted to the IEEE for possible
publication. 
Copyright may be transferred without notice, after which this version may no longer be accessible.}%
\thanks{$^{1}$Xinqi Gao is with the Department of Industrial Engineering,
        Purdue University, West Lafayette, IN 47906, USA {\tt\small gao131@purdue.edu}}%
\thanks{$^{2}$Mario Ventresca is with the Department of Industrial Engineering,
        Purdue University, West Lafayette, IN 47906, USA {\tt\small mventres@purdue.edu}}%
}

\begin{document}

\maketitle

\thispagestyle{empty}
\pagestyle{empty}

\begin{abstract}

Limited work has examined the strategic behaviors of relational networked learning agents under social dilemmas, and has overlooked the intricate social dynamics of complex systems. We address the challenge with Socio-Relational Intrinsic Motivation (SRIM), which endows agents with diverse preferences over sub-graphical social structures in order to study the impact of agents' personal preferences over their sub-graphical relations on their strategic decision-making under sequential social dilemmas. Our results in the Harvest and Cleanup environments demonstrate that preferences over different subgraph structures (degree-, clique-, and critical connection-based) lead to distinct variations in agents' reward gathering and strategic behavior: individual aggressiveness in Harvest and individual contribution effort in Cleanup. Moreover, agents with different subgraphical structural positions consistently exhibit similar strategic behavioral shifts. Our proposed BCI metric captures structural variation within the population, and the relative ordering of BCI across social preferences is consistent in Harvest and Cleanup games for the same topology, suggesting the subgraphical structural impact is robust across environments. These results provide a new lens for examining agents' behavior in social dilemmas and insight for designing effective multi-agent ecosystems composed of heterogeneous social agents.
\end{abstract}

\section{Introduction}\label{sec:intro}

Under the taxonomy examined in \cite{goforth2004topology}, $80.6\%$ of two-player games have mixed-motive structure\cite{dafoe2020open}, where individuals face the challenge of balancing whether to act selfishly or for the collective benefit. This leads to collective action problems (CAPs), where if individuals act independently and rationally, it would result in the least favorable outcome for everyone. CAPs have been widely studied in economics, psychology, sociology, and political science, and have now been extended to intelligent autonomous learning agents\cite{leibo2017multi}. Many CAPs in multi-agent social dilemmas can be classified into two broad categories: public goods dilemmas and commons dilemmas\cite{hughes2018inequity}.  Extensive research in psychology and sociology demonstrated humans sustain high level of cooperative ability across various social dilemmas. According to Interdependence Theory\cite{kelley1978interpersonal}, individuals' strategic behavior is not driven by the given extrinsic objective, but rather is transformed by the received objective into effective utility functions given the individual's social preferences over the interpersonal structures, which shape human groups' interaction. Growing research\cite{mckee2020social}\cite{madhushani2023heterogeneous} has emerged to study learning agents' behavior under Sequential Social Dilemmas(SSDs) leveraging interdependence theory's outcome transformation to construct agents' intrinsic motivation.

Two fundamental challenges in modeling autonomous agents' behavior under social dilemmas are: (1) the diversity of the population, such as individual preferences that can drive distinct decision-making; (2) the complex social structures defined by various relational and spatial constraints that shape intricate dynamics between individuals within the system. Current research simplifies the problem by assuming homogeneity and overlooks the intricate social dynamics of complex systems.

\subsection{Overview of the Proposed Solution}\label{sec:Proposal}

The model we propose endows agents with diverse preferences over sub-graphical social structures, inspired by real-world complex systems containing diverse interpersonal micro-structures. However, limited work has examined the strategic behaviors of networked mixed-motive agents under social dilemmas. This model combines graphical games, sub-graphical-based sociology, and reward-shaping multi-agent reinforcement learning (MARL) to address the aforementioned fundamental challenges, which is unexplored jointly. 

\textit{Research Question:}\label{sec:RQ} Do agent preferences over their social sub-graphical structure lead to recurring strategic patterns under Sequential Social Dilemmas (SSDs) at convergence? 

\textit{Hypothesis:}\label{sec:hypo} We hypothesize that sub-graphical structure constraints produce distinct structure-driven behaviors, leading to complex inter-group dynamics of competition and alliance that would challenge system-wide collaboration. 

We construct an incentive design model based on sub-graphical relational structures to disentangle the impact of macro- and microscopic topological structures on learning agents' strategic behavior under networked SSDs, which provides subgraphical  structurally dependent parametrized outcome transformation\cite{mckee2020social}. Here, the Socio-Relational Intrinsic Motivation (SRIM) model is a design of agents' intrinsic reward that incorporates their socio-relational structure with personal preferences. Specifically, the learning agents can have distinct heterogeneous personal preference portfolios toward neighbors across different subgraph connections. 


\textit{Contributions:}\label{sec:Structure_Contribution} To the best of our knowledge, there is no prior study that systematically investigates the impacts of various sub-graphical structures in the mixed-motive domain. This is the first work that combines graphical games, sub-graphical-based sociology, and reward shaping MARL, bridging complex systems' subgraphical structure with agents' diverse personal preferences to examine strategic behavior on mixed-motive networked learning agents. The comparison of our proposed model to existing state-of-the-art MARL models is presented in Table~\ref{tab:sota_comparison}. 

Furthermore, by carefully modeling the agents' intrinsic motivation and investigating isolated sub-graphical preferences, we disentangle the agents' distinct sub-graphical structures' impact on strategic behavior, providing structural and behavioral interpretability over complex social systems dynamics. Our novel collective bridging capacity index (BCI) is designed to examine the modification of agents' preference impact on collective decision-making. Specifically, the experimental results validate our hypothesis and reveal that variation in agents' preferences results in meaningful and distinct strategic behavioral shifts with the same underlying topology across different SSD setups. This new lens for investigating SSDs examines aggregate collective metrics and individual-level behavior, leads to more insightful future research opportunities.

\begin{table}[t]
	\caption{Comparison with state-of-the-art marl methods.}  
	\label{tab:sota_comparison}
    \centering
    \setlength{\tabcolsep}{4pt}   

	\begin{tabular}{lcccccc}\toprule
		Method & Reward & Mixed & Topology & Sub- & $\#$Agent \\
		& Shaping & Motive &  & Graph & $N$ \\ \midrule
		Haeri\cite{haeri2022reward} & \checkmark & $\times$ & \checkmark & $\times$ & 3 \\
        Hughes et al.\cite{hughes2018inequity} & \checkmark & \checkmark & $\times$ & $\times$ & 5 \\
		Jaques et al.\cite{jaques2019social}  & \checkmark & \checkmark & $\times$ & $\times$ & 5 \\
		Willis $\&$ Luck\cite{willis2023resolving} & $\times$ & \checkmark  & $\times$ & $\times$ & 2  \\
		Roesch et al.\cite{roesch2024selfishness} & \checkmark  & \checkmark & $\times$ & $\times$ & 5  \\
		Huang et al.\cite{huang2024adasociety} & \checkmark & \checkmark & \checkmark & $\times$ & 8 \\
		SRIM (proposed) & \checkmark  & \checkmark & \checkmark & \checkmark & 5,7,9,20 \\ \bottomrule
	\end{tabular}\\
\end{table}

\section{Related Work}\label{sec:Related_Works}

\textit{Sequential Social Dilemmas (SSDs):}\label{sec:SSD_intro}
The Sequential Social Dilemma model extends the Matrix Game Social Dilemma (MGSD) to capture the core decision-making process of learning agents\cite{leibo2017multi}. To resolve SSDs, numerous solutions have been proposed to motivate selfish agents to act prosocially. Designing proper intrinsic rewards to reshape agents' payoff structures, which subsequently redirects agent behavior, lies in the field of intrinsic motivation design. Previous works formulating curiosity\cite{pathak2017curiosity}, inequity aversion\cite{hughes2018inequity}, and pro-sociality\cite{peysakhovich2017prosocial} as well as causal influence assessed via
counterfactual reasoning\cite{jaques2019social}, or selfishness level guided reward shaping\cite{roesch2024selfishness} to mitigate social dilemmas and promote cooperation. These reward shaping models transformed agents' payoffs with the consideration of collective benefit, to promote prosocial behavior under social dilemmas. However, the state-of-the-art reward shaping methods lack the ability to model complex social systems, populations with heterogeneous preferences, and intricate dynamic interplay.

\textit{Networks as Relational Structures:}\label{sec:Relational_Structure} Research that utilizes networks as relational structures focuses on social relations and investigating influences without requiring direct messaging is limited in multi-agent learning systems\cite{haeri2022reward}. To overcome this, Haeri\cite{haeri2022reward} proposed the relational structure as a network framework with the consideration of social interactions between agents. This RSRN model raised the concept of `cares about' relationship structures to determine how rewards are shared between agents. Via constructing the RSRN model, emergent behaviors were examined in six different (3-agent) relational network structures, and found that different relational structures can lead to different learning agents' behaviors in the Multi-agent Particle Environment (which was examined under cooperative settings only, lacking examination under mixed-motive settings). Following this, graphical dilemmas are proposed where networks define agent-pairwise interactions only, with a minimal reward transfers derived using linear programming to optimally resolve the collective action social dilemma\cite{willis2024resolving}. Experiments are conducted in the 2-agent Harvest environment to promote cooperation\cite{willis2023resolving}. Meanwhile, utilizing networks as socio-relational structures by incorporating the social component also influences agents' strategic decision-making\cite{huang2024adasociety}.


\section{Problem Formulation}\label{sec:problem_formulation}

We extend the sequential social dilemma Markov game with the relational structure to the N-player socio-relational networked partially observable Markov game $\mathcal{M'}$, represented as a tuple $(\mathcal{S},\mathcal{A},\mathcal{O},\mathcal{T},\mathcal{R},\mathcal{G})$ defined on a finite set of states  $\mathcal{S}$, with actions $\mathcal{A}=\{\mathcal{A}^{1}, \mathcal{A}^{2},...,\mathcal{A}^{N}\}$. The observation function  $\mathcal{O}:\mathcal{S}\times \{1,2..., N\}\to \mathbb{R}^d$ specifies every agent's d-dimensional view. The joint action of the N agents $a^1, a^2, \cdot\cdot\cdot,a^N \in \mathcal{A}^{1}, \mathcal{A}^{2},...,\mathcal{A}^{N}$ leads to state changes based on a joint probabilistic transition function $\mathcal{T}: \mathcal{S}\times \mathcal{A}^{1} \times \cdot \cdot \cdot \times \mathcal{A}^{N} \to \Delta(\mathcal{S})$. The observation space is $\mathcal{O}_{i}=\{o^i\mid s\in \mathcal{S}, o^i=\mathcal{O}(s,i)\}$ for agent $i$. The agents learn and apply the policy $\pi_{i}:\mathcal{O}^{i} \to \Delta(\mathcal{A}^{i})$ to choose actions. 

Let graph $\mathcal{G}=(\mathcal{V},\mathcal{E})$ be undirected and represent the socio-relational connections of all networked agents. Here, $\mathcal{V}$$=\{{1},{2},\cdot\cdot\cdot,{N}\}$ is the non-empty set of networked agents as vertices of $\mathcal{G}$, and $\mathcal{E}$ $\subseteq \{\,(i,j)\mid i,j\in V,\;i\neq j\,\}$ represents the set of edges of $\mathcal{G}$. Agents $i,j \in \mathcal{V}$ are connected if and only if $e_{ij}=(i,j)\in \mathcal{E}$. An agent $i$ is a neighbor of an agent $j$ in graph $\mathcal{G}$ if there exists an edge $(i,j)\in \mathcal{E}$. Each agent receives its extrinsic reward defined as $ r_{\text{env}}^{i}(s_t, \vec{a}_t): \mathcal{S}\times \mathcal{A}^{1} \times \cdot \cdot \cdot \times \mathcal{A}^{N} \to \mathbb{R}$, and its socio-relational influence reward defined as $ r_{\text{socio}}^{i}(s_t, \vec{a}_t; \mathcal{G} ):\mathcal{S} \times \prod_{j\in N_{\mathcal{G}}^{pref}({i})} \mathcal{A}^{j}\to \mathbb{R} $. The preferred neighborhood portfolio for agent $i$ is defined as $N_{\mathcal{G}}^{pref}({i})$. Each agent's objective is to maximize its own long-term payoff $\mathcal{V}_i^{\overrightarrow\pi}(s_0)=\mathbb{E} \Bigg[ \sum_{t=0}^{\infty} \gamma^{t} r^{i}(s_t, \vec{a}_t) \Bigg]$, where $r^i$ is agent's total reward with temporal discount factor $\gamma^{t} \in [0,1]$.

\subsection{Preferred Neighborhood Portfolio}\label{sec:pnd} 
The traditional neighborhood for agent $i$ is $N_{\mathcal{G}}({i})=\{j\mid {(i,j)\in \mathcal{E}}\}$, which is the set of all nearest neighbors of agent $i$. Therein, the preferred neighbor concept is introduced as the Clique-neighbor and Critical-Connection neighbor. An agent has its own personal preference toward its neighbors based on how preferred the neighbor is sub-graphically connected with agent $i$, or how much the agent cares about a specific sub-graphical connection with its neighbors. 

An agent $k \in N_{\mathcal{G}}^{clique}({i})$ is a Clique-neighbor of agent $i$ if $ \exists \mathcal{C} \subseteq \mathcal{V} $ s.t. $\{i,k\} \in \mathcal{C}$, where $\mathcal{C}$ is a clique of size 3. Inside clique $\mathcal{C}$, for every distinct pair of vertices, there exists an edge. An agent $m \in N_{\mathcal{G}}^{hbn}({i})$ is a Critical-Connection neighbor of agent $i$ who lies on the shortest path toward the most strategically positioned agents $j$ with the highest betweenness centrality. The shortest-path between agent $i$ and agent $j$ is $d_{\mathcal{G}}(i,j)$.then agent $m$ $\in N_{\mathcal{G}}^{hbn}({i})$ if and  only if $d_{\mathcal{G}}(i,j)=d_{\mathcal{G}}(i,m)+d_{\mathcal{G}}(m,j)$.

The preferred neighborhood portfolio for agent $i$ is $N_{\mathcal{G}}^{pref}({i})$, which is a weighted set of all its preferred neighbors, depends on its personal perception and preferences. $\alpha, \beta, \omega$ are defined as agent $i$'s preferential weights that an individual assigns to their preferred neighbors' subgraphical structure: nearest-neighbor $N_{\mathcal{G}}({i})$, Clique-neighbor $N_{\mathcal{G}}^{clique}({i})$, and Critical-Connection-neighbor $N_{\mathcal{G}}^{hbn}({i})$. For agents who value only their direct connected neighbors,  $\alpha>0$, $\beta,\omega =0$. For agents who purely prefer social support and neighbors who have already formed a closed triangular closure, $\beta>0$, $\alpha,\omega=0$. If $\omega>0$, $\alpha,\beta=0$, agents value neighbors who are on the shortest path to the critical connection neighbor.

\subsection{Socio-Relational Influence}\label{sec:social_relational_influence}  The ego-centered socio-relational structure serves as a type of socio-relational influence as intrinsic motivation, which gives an agent extra intrinsic reward based on its personal perception and preference of its preferred neighborhood portfolio. This produces subgraphical structurally-dependent transformation and yields an effective reward structure. As a result, $
r_{\text{tot}}^{i}(s_t, \vec{a}_t; \mathcal{G} ) = r_{\text{env}}^{i}(s_t, \vec{a}_t) + r_{\text{socio}}^{i}(s_t, \vec{a}_t; \mathcal{G} )
$, where $r_{\text{env}}^{i}(s_t, \vec{a}_t)$ is the environment reward as extrinsic reward, and $ r_{\text{socio}}^{i}(s_t, \vec{a}_t; \mathcal{G} )$ is the intrinsic socio-relational influence reward based on its connection toward other agents in its preferred neighborhood portfolio. Similarly, for preferred neighborhood as Clique-neighbor only, the $ r_{\text{socio}}^{i}(s_t, \vec{a}_t; \mathcal{G} )=\beta \sum r_{\text{env}}^{k}(s_t, \vec{a}_t)$ where $k \in N_{\mathcal{G}}^{clique}({i})$, with preferred neighborhood is Critical-Connection-neighbor only, the $ r_{\text{socio}}^{i}(s_t, \vec{a}_t; \mathcal{G} )=\omega \sum r_{\text{env}}^{m}(s_t, \vec{a}_t)$ where $m \in N_{\mathcal{G}}^{hbn}({i})$. For agent $i$ with a complete preferred neighborhood portfolio, the total reward is: $ r_{\text{tot}}^{i}(s_t, \vec{a}_t; \mathcal{G})=\alpha\sum_{j\in \mathcal{N}_{\mathcal{G}}}r_{\text{env}}^{j}(s_t, \vec{a}_t) + $ $\beta\sum_{k\in \mathcal{N}_{\mathcal{G}}^{clique}}r_{\text{env}}^{k}(s_t, \vec{a}_t)  + \omega\sum_{m\in \mathcal{N}_{\mathcal{G}}^{hbn}}r_{\text{env}}^{m}(s_t, \vec{a}_t)+r_{\text{tot}}^{i}(s_t, \vec{a}_t)$. The subgraphical-dependent weights $\theta=(\alpha, \beta,\omega)$ can set to be one-hot, mixed, or dynamically scheduled.


\section{ Environmental Setup}\label{sec:Env_Section}

To investigate the impact of subgraphical structures on learning agents' strategic behavior under networked SSDs, we conduct experiments on two benchmark sequential social dilemma environments: a common-pool resource game Harvest, and a public goods game Cleanup\cite{leibo2017multi}\cite{hughes2018inequity}.

In Harvest, each agent's goal is to compete and collect apples that replenish according to a spatially dependent regrowth rate. Among the 8 actions agents can make, ``FIRE'' is a punishment action against other agents that costs the agent one reward unit, while being hit by the beam cost 50 reward units. In Cleanup, agent's goal is to collect apples. The apples respawn rate controlled by a separate aquifer that accumulate waste over time, linearly reducing apple respawn rate until spawning stops beyond a saturation threshold. Agent has an additional ``CLEAN'' beam action that removes waste from the aquifer, enabling apples sustainable regeneration.

\textit{Experiment Metrics:}\label{sec:experiment_metrics} The experimental metrics are used to test the hypothesis that sub-graphical structures constraints produce distinct structure-driven behaviors under SSDs (rather than to resolve social dilemmas.) We design collective networked related metrics and investigate individual-level analyses as the primary objective.

\subsubsection{Structural Evaluation}\label{sec:BCI_def}
The Bridging Capacity Index (BCI) is proposed to reflect how well agents strategically obtain and carry resources collectively in a complete episode, given that agents have different capabilities of bridging disconnected relational substructures. It is a weighted aggregation that normalizes individual gains by their varying capabilities, constrained by the relational structure, and reflects how much they contribute compared to the overall collective resources at convergence. Consequently, we can identify strategic shifts in learning agents under different agent preferences across various topologies at the system level, which provides a foundation for subsequent individual-level investigation. 
Classical network centralities focus solely on topological structure and offer no insight into individual or collective resource collection. BCI quantifies who gained environmental rewards given their structural brokerage opportunities and is sensitive to reward shifts across nodes. Hence, it is applicable to any networked MARL setting with per-agent rewards and provides a population-level indicator of reward shifts across individuals before conducting mesoscopic-level behavioral analyses.

Consider the case with $N$ networked agents in SSDs. Agent $i$ collects apples as an extrinsic environmental reward,  $\{R^i_{env}=\sum_{t=1}^Tr_{\text{env},\,t}^{i}\mid 1,2,...T\}$, where $T=1000$ for a complete training episode. The BCI metric is then defined as: 
\begin{align}
BCI= \frac{\sum_{i=1}^N(1-C_{i})R^i_{env}}{\sum_{i=1}^N R^i_{env}}  
\end{align}
where $C_{i} = \sum_{j\in N_i}c_{ij}$ is Burt's constraint for node $i$ and is a measure of redundancy\cite{burt2018structural}, with $c_{ij}=(p_{ij}+\sum_{q\in N_i, q \neq i,j} p_{iq}p_{qj})^2$ is the dyadic constraint of node $i$ with respect to $j$. Here $p_{ij}$ quantifies the normalized direct share of ego node $i$'s total connection that goes to node $j$, $\sum_{q\in N_i, q \neq i,j}p_{iq}p_{qj}$ represents the aggregated indirect investment share of node $i$ in $j$ through their mutual contacts $q$. For a given graph,  $p_{ij}=\frac{a_{ij}+a_{ji}}{\sum_{j}(a_{ij}+a_{ji})}$ where $a_{ij}\in \{0,1\}$ with $a_{ij}=1$ if and only if $(i,j)\in \mathcal{E.}\ $ Higher redundancy of node $i$ increases $c_{ij}$ and cumulatively increases $C_i$. Hence, the complement of Burt's constraint $1-C_{i}$ quantifies a node's bridging capability and measures the extent to which a node serves as a bridge between two disconnected groups. Higher $1-C_{i}$ indicates the node has higher bridging capacity and spans more structural holes. BCI metric is a reward-weighted average of agents' bridging capacities. The higher BCI, the more resources are collectively held by agents with higher bridging capacity in networked systems.

Mathematically, BCI couples topological structures with environmental rewards. It utilizes individual-level structural properties to form an aggregated reward-weighted average metric that is invariant to reward scaling and satisfies the monotonicity property with respect to the collection of environmental rewards for higher bridging structural agents. BCI is scale-invariant in rewards, but designed to be sensitive to relative shifts in reward obtained across agents, reflecting the structural reward distribution at the population level. Furthermore, each topological structure inherently determines the topology-constrained theoretical range (TCTR) for the bridging capacity. Since BCI is a convex combination of ${1-C_i}$ when $R^i_{env} \geq 0$, the topology-constrained bound is $[\min_{i\in \mathcal{V}}{(1-C_i)},\max_{i\in \mathcal{V}}{(1-C_i)}]$. The actual BCI metric collected reflects the true bridging capacity.

\subsubsection{Individual Evaluation}\label{sec:ind_measure}
(i) The base reward $R_{env}^i$ is the actual extrinsic reward each agent gathered in one episode. (ii) Individual aggressiveness is the number of times an agent chooses the ``FIRE'' action within an episode and enables us to examine the emergence of distinct aggressiveness in each agent. (iii) Individual public good contribution effort ($E^i)$ is the number of times an agent choose the  ``CLEAN'' action and make effort to reduce waste in the aquifer. This enables us to examine the emergence of prosocialness in agents in Cleanup. Further, we constructed a [0,1] bounded metric Social Contribution Index $SCI=\frac{E^i+\epsilon}{E^i +A^i +2\epsilon} $ where $E^i$ is the effort contributed and $A^i$ is the total apple collected in one episode to measure the individual trade-off between prosocial behavior and self-interested behavior, with 0 indicating the agent is self-interested and acts as a pure harvester, 1 indicating the agent is prosocial and acts as a pure cleaner.

\section{Experimental Setup and Results}\label{sec:Exp_Section}

\begin{figure}[t]
    \centering

    \includegraphics[width=0.8\columnwidth]{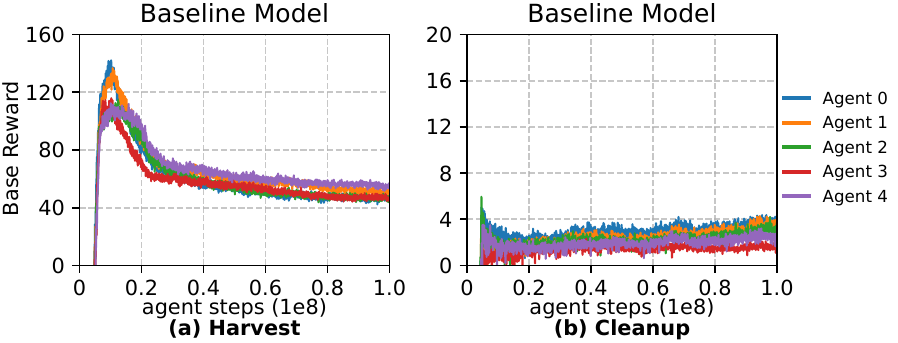}
    
    \caption{Individual base reward under the baseline model exhibits identical decaying trajectories. 5 agents with network-free and preference-free protocol learn and compete independently under Harvest and Cleanup.}
    \label{fig:inv_BR_baseline}

\end{figure}

The goal of the experiments is to examine whether the addition of the MARL agents' socio-relational influence based on sub-graphical structure has a significant impact on agent resource collection capability and strategic behavior. 

Our experiments utilize the default baseline neural network architecture\cite{hughes2018inequity} and train agents using the Proximal Policy Optimization (PPO) algorithm, which collects experiences from 20 parallel and independent workers. Each worker runs a distinct environment instantiation with all agents playing simultaneously, generating agents' reward and metrics data that are aggregated synchronously into training batches. The baseline for all experiment scenarios is the network-free and preference-free protocol to ensure fair comparisons. Figure \ref{fig:inv_BR_baseline} presents the individual base reward trajectories under the baseline models, at $1e8$ timesteps the mean base reward is 48.31 in Harvest and 3.02 in Cleanup.

\textit{Experiment Setup:} A series of experiments is designed across seven different 5-agent networks and three customized multi-agent network configurations with increasing population sizes of 7, 9, 20. The analyzed networks are shown in Figure~\ref{fig:basereward_indiv}; full details in the Appendix.\footnote{Detailed topologies examined N=5,7,9,20, further statistical details, and model robustness analysis are in our Appendix Section III} Fig 1. The networks are selected to capture key sub-graphical structure variations, including degree-$\alpha$, clique-$\beta$, and bridge-$\omega$ structures. To disentangle the sub-graphical structure impact and identify distinct sub-graphical-driven behaviors, the individually activated sub-graphical preferences are systematically examined across a spectrum of topologies, providing sound interpretability for complex situations where agents hold mixed preferences across different sub-graphical structures.

\subsection{Experimental Results: Structural Evaluation}

\begin{table}[t]
\caption{Bridging capacity across variations over 5-agent network configurations under traditional neighbor preference. }
\label{tab:bci-table}
\begin{center}
\begin{minipage}{\columnwidth}
\centering
\setlength{\tabcolsep}{4pt}  
\begin{tabular}{l*{5}{c}} 
\toprule
Topology & $Med_H$ & $IQR_H$$^{a}$ & $Med_C$ & $IQR_C$ & LB-UB$^{b}$ \\
\midrule
Bipartite & 0.525  & 0.520-0.529 &  0.501  & 0.500-0.504 & 0.500-0.667 \\
Complete  & 0.234  & 0.234-0.234& 0.234  & 0.234-0.234 & 0.234-0.234 \\
Cycle     & 0.500  & 0.500-0.500 & 0.500  & 0.500-0.500 & 0.500-0.500 \\
House     & 0.378  & 0.368-0.380 & 0.331  & 0.315-0.349 & 0.111-0.500 \\
Star      & 0.023  & 0.022-0.032 & 0  & 0-$10^{-10}$ & 0.000-0.750 \\
Symmetric & 0.228  & 0.226-0.230 & 0.220  & 0.219-0.224 & 0.219-0.316 \\
Wheel     & 0.339  & 0.338-0.340 & 0.343 & 0.340-0.344 & 0.306-0.344 \\
\bottomrule
\end{tabular}
\par\vspace{2pt}
\raggedright
{\footnotesize $^{a}$  IQR: Q1-Q3 range, with Q1: The 1st quartile; Q3: the 3rd quartile. H:Harvest, C:Cleanup}
{\footnotesize $^{b}$ Topology-constrained theoretical lower and upper bounds are shown as LB and UB.}
{\footnotesize $^{c}$ Results across over 5 random seeds}\\
\end{minipage}
\end{center}
\end{table}


\textit{BCI under Nearest Neighbor Preference:} A comparison of the BCI statistic across different topologies is performed to evaluate whether the total resource collectively carried by agents with higher bridging capacity occurs under a specific topology at convergence. It is observed that distinct BCI values vary across different networked SSD systems for both Harvest and Cleanup environments. Further, topologies with higher BCI values indicate that more resources are collectively held by agents with higher bridging capacity. Table \ref{tab:bci-table} shows the actual bridging capacity of the networked systems represents how well agents with structurally higher bridging capacity strategically attain resources compared to the overall collective resources at convergence  under nearest neighbor preferences across each 5-agent network configuration.

\begin{figure}[t]
    \centering
    \includegraphics[width=0.95\columnwidth]{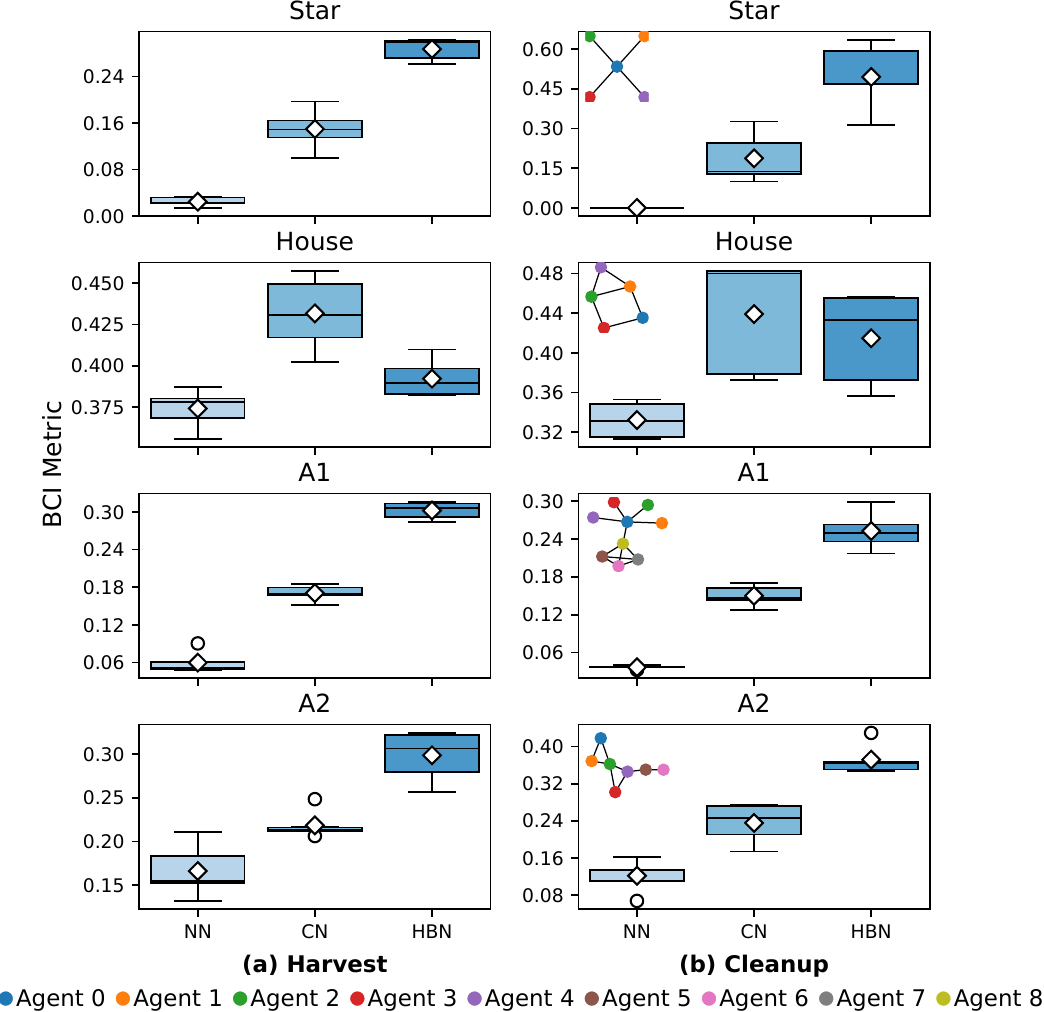}
    
    \caption{\label{fig:bci-diff_pref} The distribution of BCI for various topologies under different agent social preferences under Harvest and Cleanup. The y-axis represents the BCI metric. `NN': Nearest-Neighbor preferences, `CN': Clique-Neighbor preferences, `HBN': Critical-Connection Neighbor preferences. Topologies are present in each subplot. Agents' colors are consistent across all subfigures.}

\end{figure}

\textit{BCI under Different Agent Preferences:}
The BCI metric across different types of social preferences was evaluated while keeping the underlying network topology constant for both Harvest and Cleanup. As mentioned in Section~\ref{sec:Env_Section}, although each networked system has its own theoretical range, the empirical BCI reflects the true bridging capacity. Figure \ref{fig:bci-diff_pref} shows that modifications to agent preferences over sub-graphical structures lead to distinct stabilized BCI metrics distributions at convergence under the same network configuration under both SSD systems. This indicates that varying agents' preferences can lead to a shift of resources collected strategically by agents with different levels of bridging capacity, resulting in systematic redistribution of reward across different structural positions in the population under the same topology within the same SSD environment.

Furthermore, by comparing the distributions of distinct BCI metrics under different agent preferences in the Harvest and Cleanup, with the same topology fixed (tested with $N=5,7,9,20$), it is found that the relative ordering of the BCI distributions across preferences is preserved across these two SSD environments. This indicates that each subgraphical preference consistently yields a similar structural reward distribution across the population. Our proposed BCI metric successfully captures the impact of strategic behavioral structural shifts within the population and has been shown to be generalizable across different networked SSD systems.

Specifically, under the critical-connection neighbor preference, the agent with the highest betweenness centrality has been cared the most and obtains the highest resources under both Harvest and Cleanup, resulting in the highest mean BCI value in the Star, A1, A2 topologies, compared to scenarios with agents holding preference over other sub-graphical structures under the same topologies. Meanwhile, under the nearest neighbor preference, the agent with the highest degree centrality cares for others the most, hence acting most prosocially with the least reward collected, which leads to the lowest mean BCI value in the Star, House, A1 and A2. The BCI metric reveals the shift in rewards with respect to structurally advantageous positions, hence provides a population-level indicator of shifts in reward distribution across individuals before conducting individual-level analysis. These distinct BCI metrics reflect the significance of studying how strategic behavior changes and equilibrium shifts with agents holding different social preferences. Given these results, individual evaluations to further understand individual-level behavioral variations were then conducted.

\subsection{Empirical Results: Individual Evaluations}\label{sec:indiv_eval}

\begin{figure*}[t]
    \centering
    \includegraphics[width=1\textwidth]{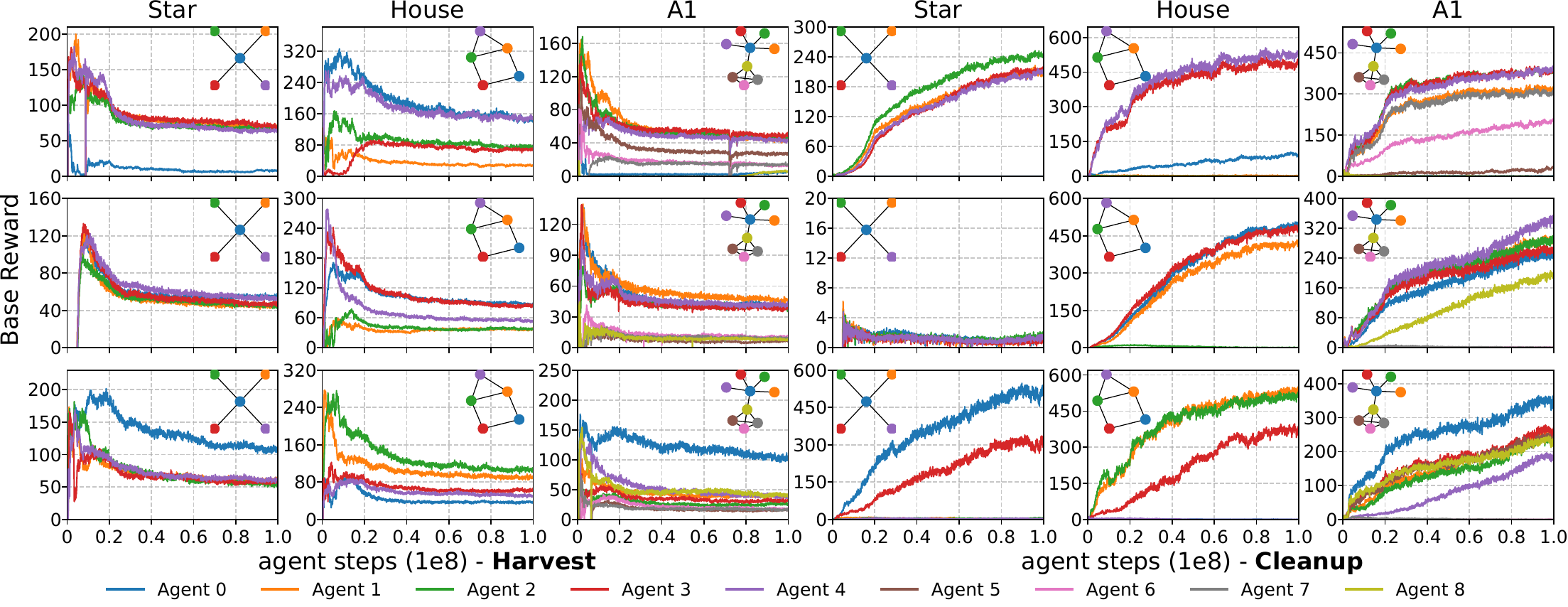}
  
    \caption{\label{fig:basereward_indiv} Individual mean base reward under different agent preferences across different topologies. Top row: Nearest Neighbor Preference ($\alpha=1$); Middle row: Clique-Neighbor Preference ($\beta=1$); Bottom row: Critical-Connection Neighbor Preference ($\omega=1$). Non-active parameters set to 0. Topologies are shown in the top-right (top-left) corner of each subplot. Star, House has 5 agents; A1 has 9 agents; The agents' colors are consistent across all subfigures. A2 (intermediate size N=7) shows similar trajectories across preferences; its visualization (same for Fig.5) is therefore deferred to App.Sec.III for readability.}

\end{figure*}

\subsubsection{Base Reward under Different Agent Preferences}\label{basereward_sec} An analysis of the individual base reward trajectory under three distinct agent preferences over sub-graphical structures was performed under both Harvest and Cleanup, each evaluated across different topologies with agent preference held invariant, as illustrated in Figure~\ref{fig:basereward_indiv}. Under the Nearest Neighbor Preference, a pattern has been observed in most cases in both Harvest and Cleanup: among the agent who cares for its neighbors the most are the most prosocial agent, at least one agent with the highest degree centrality obtains the lowest resources across different topologies; while agents who are most self-interested prioritize their own immediate benefits over the long-term collective benefit and over-exploit the resources the most (Harvest) or free-ride on other agents' efforts on cleaning benefit (Cleanup). Under Clique Neighbor Preference, a different recurring pattern has been revealed, where agents who form clique alliances are constrained by clique preference and are worse off in gathering resources compared to agents outside the alliances. The agents without sub-graphical constraint are able to act selfishly, exploiting the prosocial agents and gathering relatively higher resources in Harvest and Cleanup. In contrast, the individual base reward under Critical-Connection Neighbor Preference exhibits a pattern where agents with the highest betweenness benefit from this topological position and preference, achieving the highest individual resource collection in both SSDs systems.

These empirical findings confirm our hypothesis: even when the underlying topology remains constant, and the apple regrowth rate remains as default, agents holding different preference portfolios over sub-graphical structures produce distinct and observable shifts in resource gathering patterns in both SSDs systems. This motivates into further investigation over individual aggressiveness in Harvest, and individual public goods contribution effort in Cleanup.

\begin{table*}[!t]
\caption{Mean Summed individual aggressiveness under nearest neighbor social preference for each stage across five seeds. }
\label{tab:combined_ind_aggressiveness}
\begin{minipage}{\textwidth}
\centering
\begin{tabular}{@{}l *{15}{c}@{}}
\toprule
 & \multicolumn{5}{c}{1st Stage$^{a}$: After Exploration Trials} & \multicolumn{5}{c}{2nd Stage: Transient Stage} & \multicolumn{5}{c}{3rd Stage: At Convergence} \\
\cmidrule(lr){2-6} \cmidrule(lr){7-11} \cmidrule(lr){12-16}

\multirow{-2}{*}{Agent-id} & 0 & 1 & 2 & 3 & 4 & 0 & 1 & 2 & 3 & 4 & 0 & 1 & 2 & 3 & 4 \\
\midrule
Wheel & \textbf{1.23$^{b}$} & 2.73 & 2.57 & 3.56 & 2.30 & \textbf{46.40} & 29.85 & 22.74 & 21.14 & 28.75 & \textbf{16.56} & 5.26 & 7.76 & 2.79 & 13.45 \\
House & 3.58 & \textbf{1.99} & \textbf{2.13} & 3.57 & 5.73 &  40.13 & \textbf{77.83} & \textbf{37.93} & 39.11 & 75.06 & 5.18 & \textbf{20.07} & \textbf{19.70} & 6.45 & 17.52 \\
Bipartite& \textbf{1.37} & \textbf{0.87} & 8.48 & 3.01 & 3.55 & \textbf{220.81} & \textbf{120.42} & 49.12 & 28.74 & 47.68 & \textbf{6.21} & \textbf{8.35} & 4.99 & 3.65 & 4.84 \\
Symmetric &  \textbf{0.57} & \textbf{0.61} & 4.31 & 3.89 & 2.62 & \textbf{114.04} & \textbf{211.06} & 36.39 & 31.70 & 54.69 & \textbf{19.47} & \textbf{9.37} & 9.89 & 3.93 & 8.55\\
Star & \textbf{1.22} & 6.71 & 34.20 & 40.62 & 10.34 & \textbf{503.17} & 128.39 & 146.90 & 110.74 & 82.79 & \textbf{18.51} & 17.46 & 24.52 & 22.07 & 13.33 \\
\bottomrule
\end{tabular}
\par\vspace{0.01pt}
\raggedright
{\footnotesize $^{a}$ 1st Stage: Post-random exploration phase that beginning at $10^6$ timesteps. The 2nd Stage transient phase: Phase Following 1st Stage to $6 \times 10^7$ timesteps, characterized by a sharp increase in aggressiveness of the agents with the highest degree centrality. The 3rd Stage: Phase where the networked system reaches convergence in the last $40\%$ of the complete trained agents' steps.}
{\footnotesize $^{b}$ \textbf{Bold}: Agents with the highest degree centrality.}\\

\end{minipage}
\end{table*}

\subsubsection{Individual Aggressiveness} \textit{under Traditional Neighbor Preference:}
Given the distinct agent resource gathering patterns identified in individual base reward analysis, we examine the individual aggressiveness to further investigate agent's strategic behavior and whether the aggressiveness should be classified as cooperative or non-cooperative under SSDs. First, we examine aggressiveness under nearest neighbor preference. Table~\ref{tab:combined_ind_aggressiveness} shows that agents with the highest degree centrality tend to learn to be the least or second-least aggressive agents in the first stage of post-random exploration across 5 seeds. As learning proceeds, most of these agents go through a transient stage characterized by a sharp increase in the use of the ``FIRE'' action, followed by readjusting their aggressiveness to a lower level, but still surpassing other agents who do not have the highest degree centrality across seeds. When the game reaches convergence, most agents with the highest degree centrality maintain a higher level of mean summed aggressiveness, while gathering the least rewards in Harvest, except for the Star topology. 

Further investigation into agents' aggressiveness, individual base reward, and the socio-relational influence given agents' preference over sub-graphical structure revealed that the high level of aggressiveness does not directly indicate agents' holding a defection or cooperation policy. In each 5-agent network configuration, at least one agent with the highest degree centrality kept its mean summed aggressiveness at the highest or second-highest level among all agents and consistently ranked within the top-2 in every seed, except the Star topology. Those agents' high aggressiveness is a cooperative policy, since they use it to protect the cared group's resources from depletion. This selective sanction is imposed on agents who deplete resources, which leads to their own base reward being compromised while protecting their community's resources. Meanwhile, the agent with the highest degree centrality in Star successfully learnt this complicated cooperative policy of sanction in only 2 of 5 seeds. In the other 3 seeds, failure to learn the sanction policy with the lowest aggressiveness leads to agents on the periphery maintaining a higher level of aggressiveness and using ``FIRE'' indiscriminately and selfishly to compete against each other and obtain the immediate reward, resulting in the lowest collective return compared to all other topologies. These periphery agents' high aggressiveness is acted out of selfishness and classified as a defection policy.

\begin{figure}[t]
    \centering

    \includegraphics[width=0.95\columnwidth]{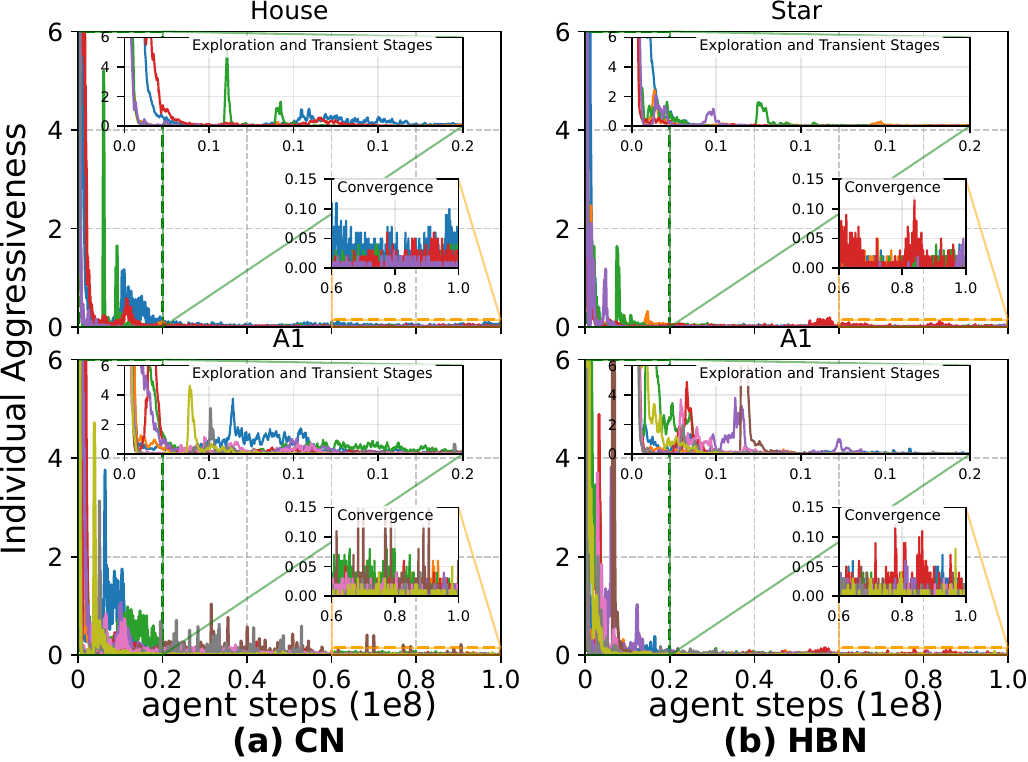}
    
    \caption{This figure shows overall agents' individual aggressiveness over the entire experimental agent steps under (a) clique neighbor preference (CN) in 5-agent House and 9-agent A1 topology, and (b) critical-connection preference (HBN) in 5-agent Star and 9-agent A1 topology. The upper subregion subfigure shows agents' individual aggressiveness in random exploration and transient stages with x-axis limit at $2e7$ and original y-scale. This lower subregion subfigure shows agents' individual aggressiveness at convergence, with the x-axis range: $[0.6e8,1e8]$, and y-axis range: $[0,0.15]$.}
    \label{fig:inv_fire_count_CN_highlight}

\end{figure}

\textit{Aggressiveness under Clique Neighbor Preference.} Figure~\ref{fig:inv_fire_count_CN_highlight}a) shows a typical result that the agents who form clique alliances and care about the clique structure preference (Agent 1, 2, 4 in House topology) learn to reduce their aggressiveness the fastest in the first exploration stage across 5 seeds, in contrast to agents outside the clique alliance (Agent 0, 3).  As learning proceeds, some agents in clique alliances (Agent 2) undergo a transient stage characterized by a sudden increase in the frequency of applying the ``FIRE'' action, followed by readjusting their aggressiveness to a lower level. This transient behavior indicates that Agent 2 strategically alters its non-aggressive prosocial behavioral policy learned in the first stage and explores the increased utilization of aggressiveness, while striving to maximize their objectives of the clique alliance benefit. At convergence, within the clique alliance, some agent (Agent 2) consistently exhibited aggressiveness typically $3\times$ higher than that of other clique alliance agents (Agent 1, 4) for House, A1 and A2 across 5 seeds, with rare cases ranging from $2\times$ to $5\times$ high. This higher aggressiveness is compatible with their clique-alliance group objectives and also costly to their own benefit, which indicates this aggressiveness as an altruistic punishment policy, whereas the other alliance agents (Agents 1, 4) failed to learn this complicated sanctioning mechanism and kept the lowest aggressiveness as their prosocial strategy. In contrast, agents outside the clique alliances (Agent 0, 3) exhibit a consistently higher level of aggressiveness compared to the agents who formed the clique alliances over the entire experimental period across seeds. This finding aligns with the base reward trajectory, where agents outside the clique structure gather more resources due to their selfish exploitation and their higher level of competition over resources. While we focus here on the analysis of the House topology, the core finding and insight generalize to A1, A2 with clique structures.

\textit{Aggressiveness under Critical-Connection Neighbor Preference.} Figure~\ref{fig:inv_fire_count_CN_highlight}b) shows a typical result for agents holding preference over the critical connection neighbor in the 5-agent and 9-agent Harvest game. Under the Star and Wheel structure, a pattern can be observed where the hub agent (Agent 0) with the highest betweenness value learns to reduce aggressiveness the slowest compared to the agents on the periphery during the early exploration phase (across 5 seeds), displayed in the upper subregion of the upper Figure \ref{fig:inv_fire_count_CN_highlight}b). This behavior is due to their strategic position being preferred by the population. As a result, Agent 0 maintains a low level of aggressiveness at the convergence. A similar pattern is observed in 7-agent A2 topology. In contrast, in A1, with further increase of population, agents with the hub position (Agent 0) and agents with a broker position (Agent 8) acting as the critical connector exhibit consistent moderate aggressiveness across 5 seeds. Combined with the individual base reward, the top two resource collectors function as critical connectors of substructures. We hypothesize that those agents with the highest betweenness centrality exhibit moderate aggressiveness relative to others, which may be due to resource scarcity constraint when the population number surpasses the environmental capacity of maintaining sustainable resources. Moreover, the investigation of individual aggressiveness under the same topology shows that changes in agents' preferences over the sub-graphical structure produce observable strategic behavioral variations. 
\subsubsection{Individual Contribution Effort} 
\begin{figure}[t]
    \centering
     \includegraphics[width=1\columnwidth]{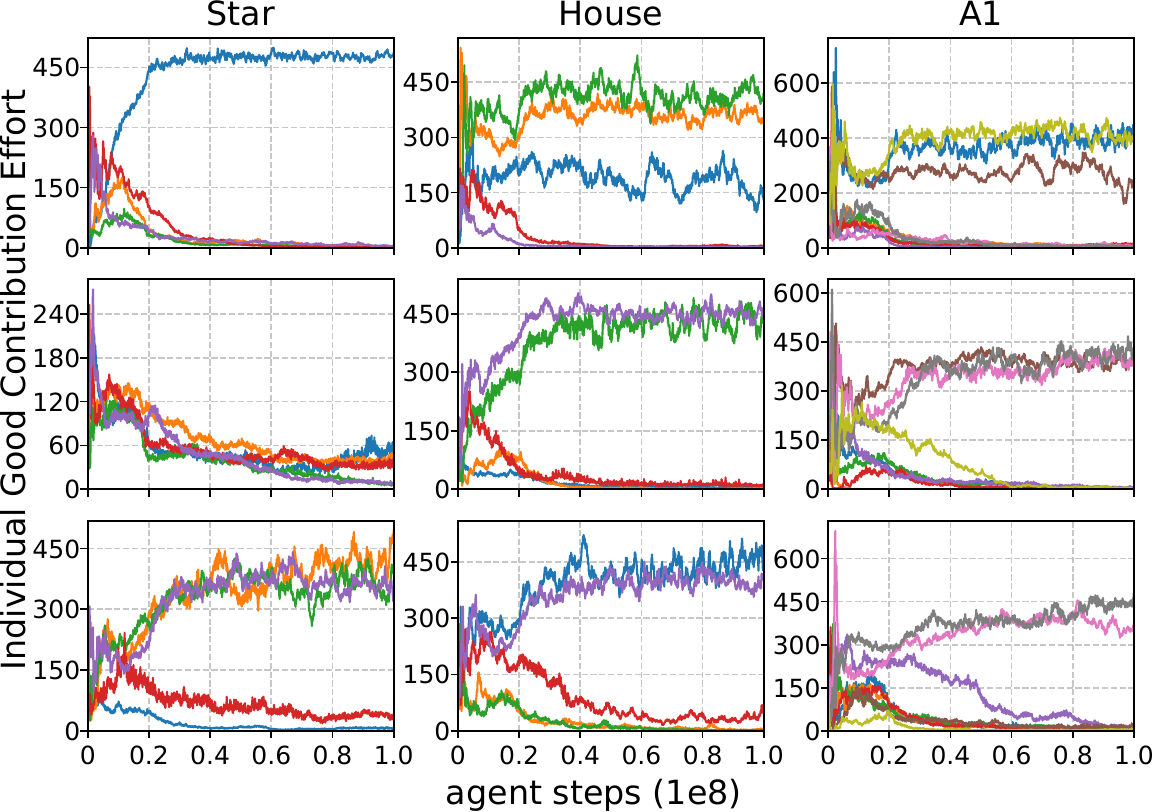}
    \caption{\label{fig:Clean_Count} Individual Good Contribution under different agent preferences across different topologies in Cleanup. Top row: Nearest Neighbor Preference; Middle row: Clique-Neighbor Preference; Bottom row: Critical-Connection Neighbor Preference. Colors are consistent with Fig.3 legend.}

\end{figure}

Given the agent apple gathering patterns identified in subsection~\ref{basereward_sec} in Cleanup, we then examine the individual public good contribution effort agents made to reduce waste to further investigate the impact of macro- and microscopic sub-graphical structures on agents' strategic behavior. Figure~\ref{fig:Clean_Count} exhibits structurally identical cleaning patterns consistently for agents that keep the same subgraphical preference across different topologies. Meanwhile, distinct cleaning patterns are observed for agents with different preferences, while keeping the topology fixed.

\textit{SCI under Nearest Neighbor Preference.} Jointly examine Figure~\ref{fig:Clean_Count} and SCI metric reveals a consistent pattern across all examined topologies: at least one of the agents with the highest degree centrality always converges to the primary cleaner role, who makes the highest individual contribution effort, meanwhile consistently gathers the least reward under the nearest neighbor preference. In topologies with only a single agent having the highest degree centrality, the agent consistently converges to this primary cleaner role. In Star topology, the hub agent (Agent 0) achieves $SCI=1$ across all 5 seeds. 
The mean SCI of Agent 0 in A1, Agent 2 in A2 is 0.992 and 0.998 (See App. Sec.III for confidence intervals). House topology with two agents (Agent 1, 2) with the highest degree centrality shows an alternating pattern, where at least one of the agents with the highest degree centrality serves as the cleaner, with mean SCI is 0.99 while the other agent serves as a cleaner or pure harvester across 5 random seeds. This indicate the agents with the highest degree centrality acted out of pure prosocial behavior, sacrificing their own benefit for reward and serving as a cleaner to clean up the waste consistently across all topologies.

\textit{SCI under Clique Neighbor Preference.} Similarly, joint analysis of Figure~\ref{fig:Clean_Count} and the SCI metric reveals that role specialization and the division of labor into pure cleaners and pure harvesters occur within agents who form clique alliances and care about clique preference across all examined topologies. Inside the clique alliance, cleaners (Agent 2, 4 in House, Agent 1, 2, 4 in A2, Agent 5-7 in A1) make the highest individual contribution effort to keep resources sustainable, and act out of prosocialness toward the cared group with mean SCI ranging from $0.983$ to $0.999$ across all topologies; with harvesters (Agent 1 in House, Agent 0,3 in A2, Agent 8 in A1) learned to decrease cleaning waste over time and turn to dedicated harvesting apples for the cared group with mean SCI ranging from $0.007$ to $0.166$ across all topologies; while agents outside of the clique alliances learnt to be pure free-rider and harvesting apples with no individual contribution effort being made, and taken advantages of the agents who were cleaning waste all the time. The cleaner role specialization consistently emerges across 5 random seeds, with each clique alliance having at least one agent dedicated to being a pure cleaner, consistently only from the clique alliance group. We observe that the specific agent learns to be dedicated to which role varies across 5 seeds. Agents with no clique structure in the topologies (Star) randomly adopt the cleaner role across 5 seeds with no consistent pattern and role specialization observed with a mean SCI of 0.668.

\textit{SCI under Critical-Connection Neighbor Preference.} Likewise, joint analysis of Figure~\ref{fig:Clean_Count} and the SCI metric reveals that the agent holding the critical connection neighbor preference with the highest betweenness value learns to always act out of self-interest and as a pure harvester and free rider, and makes the least individual contribution effort across all examined topologies under Cleanup. This is due to the strategic position being preferred by the population, enabling the agent to free-ride while other agents randomly adopt the cleaner roles to ensure the critical connection agent harvests. More specifically, Agents 0 in Star, 1, 2 in House, 2, 4 in A2, and 0, 8 in A1 have the highest and second-highest betweenness centrality value. Across all topologies, the highest betweenness agents exhibit consistently near-zero SCI, with mean SCI ranging from $0.007$ to $0.032$. This reflects that agents with the highest betweenness centrality acted purely selfishly and consistently served a harvester role across all topologies across 5 seeds. 

The SCI analysis shows that the individual trade-off between its prosocial behavior and self-interested behavior is consistent and recurring across all topologies under each subgraphical preference in Cleanup. Meanwhile, under the same topology, changes in agents' preferences over the subgraphical structure lead to observable consistent variations in agents' strategic behavior under social dilemmas.

\section{Conclusion}
In this paper, we constructed the SRIM model, which produces a subgraphical structurally dependent transformation to construct agents' intrinsic motivation. Our empirical results revealed that considering socio-relational sub-graphical preferences yields valuable insights. The change in agents' preference portfolios results in consistent distinct shifts in resource-gathering patterns, leading to distinct meaningful strategic behavioral variations across two benchmark social dilemmas (Harvest and Cleanup). This validates our hypothesis that sub-graphical structure constraint produces distinct structurally driven behaviors, confirming our reward-shaping model is robust and environmentally invariant.

\bibliographystyle{IEEEtran}
\bibliography{sample}

\appendix


The supplementary materials contain preliminaries, additional environment and experimental setups, and detailed analyses that cannot fit into the main manuscript.

\section{Preliminaries}\label{sec:intro2}

 Many CAPs in multi-agent social dilemmas can be classified into two broad categories: public goods dilemmas and commons dilemmas. Common access resources like groundwater resources, fisheries, forests, and public roads can be exploited by selfish usage of individuals, which leads to water shortage, overfishing, deforestation, and traffic congestion for society. Since common goods are rival goods but also non-excludable, this leads to the Tragedy of the Commons, where individuals act rationally but ultimately deplete the commonly shared resource. Meanwhile, public goods are non-rival and non-excludable, such as building communal irrigation systems or mitigating climate change, where costly individual contributions are incurred to generate a shared collective benefit, but each individual has an incentive to free-ride and act rationally rather than contribute.

\subsection{Sequential Social Dilemmas}\label{sec:SSDs}

A sequential social dilemma is defined as an N-player partially observable Markov game $\mathcal{M}$, represented as a tuple $(\mathcal{S},\mathcal{A},\mathcal{O},\mathcal{T},\mathcal{R})$ defined on a finite set of states $\mathcal{S}=\{\mathcal{S}_{1},\mathcal{S}_{2},..., \mathcal{S}_{N}\}$, with actions $\mathcal{A}=\{\mathcal{A}^{1}, \mathcal{A}^{2},...,\mathcal{A}^{N}\}$. The observation function  $\mathcal{O}:\mathcal{S}\times \{1,2..., N\}\to \mathbb{R}^d$ specifies every agent's d-dimensional view. The joint action of all the N agents $a^1, a^2, \cdot\cdot\cdot,a^N \in \mathcal{A}^{1}, \mathcal{A}^{2},...,\mathcal{A}^{N}$ leads to state changes based on a joint probabilistic transition function $\mathcal{T}: \mathcal{S}\times \mathcal{A}^{1} \times \cdot \cdot \cdot \times \mathcal{A}^{N} \to \Delta(\mathcal{S})$. The observation space is $\mathcal{O}_{i}=\{o^i\mid s\in \mathcal{S}, o^i=\mathcal{O}(s,i)\}$ for agent $i$. The agents learn and apply the policy $\pi_{i}:\mathcal{O}^{i} \to \Delta(\mathcal{A}^{i})$ to choose actions. The reward function is $\mathcal{R}= \{r^1, r^2, ..., r^N\}$, where for player $i$ the reward it receives is $r^i: \mathcal{S} \times \mathcal{A}^{1}, \mathcal{A}^{2},...,\mathcal{A}^{N} \to \mathbb{R} $. For mathmatical simplicity, we write $\vec{a}=(a^1,a^2, ..., a^N), $ $r^i(s,a^1,a^2, ..., a^N)=r^i(s,\vec{a})$. Each agent's objective is to maximize its own long-term payoff $\mathcal{V}^{\overrightarrow\pi}(s_0)=\mathbb{E} \Bigg[ \sum_{t=0}^{\infty} \gamma^{t} r^{i}(s_t, \vec{a}_t) \Bigg]$, where $\gamma\in[0,1]$ is the temporal discount factor.

\subsection{Structural Hole and Burt's Constraint}\label{sec:burts}

Structural holes is an important social structure in network analysis. A structural hole is proposed in \cite{burt2018structural}. It is defined as the structure where there exists a lack of connection (gap) between two groups with valuable complementary resources. This lack of connection is considered to have a brokerage opportunity for future individuals who later connect the gap, which would lead to higher advantages by locating this topological position. In a social network, a person in this social structure, a broker,  would be able to access multiple valuable and distinct information and serve as an important information flow or influence actor in the social network.

In order to examine and identify which individual connects the gap and spans the structural hole in a specific network, \cite{burt2018structural} proposed Burt's Constraint. An individual with high Burt's constraint indicates the individual has neighbors closely connected, hence the node position the individual holds has high redundant connections and lacks brokerage opportunities. Similarly, an individual with low Burt's constraint indicates the individual is in a position that spans one or more structural holes and has connections in distinct groups, hence has high bridging capacity.

The Burt's constraint is: $C_i = \sum_{j\in N_i}c_{ij}$ where

\begin{align}
c_{ij}=(p_{ij}+\sum_{q\in N_i, q \neq i,j} p_{iq}p_{qj})^2
\end{align}

Where $c_{ij}$ is the Burt's constraint, $N_i$ is the neighbors of node $i$, $p_{ij}$ is the proportion of interactions that node $i$ invested in node $j$.

\section{Experimental Setup Details}\label{sec:HarvestCleanup_Env_Section}

In the Harvest game, each agent's goal is to collect apples in a grid-world environment. The apples are the public shared resource, and regrow at a rate that varies based on the current spatial distribution of the existing apples that have not yet been collected. In an area where all the agents come and compete and collect the apples within that field within a certain time, the apples in that area would be completely depleted and no longer have the capability of regrowth. The game would reset after each episode ends, which is 1000 steps for all the agents. 

\begin{figure*}
    \centering
    \includegraphics[width=0.3\columnwidth]{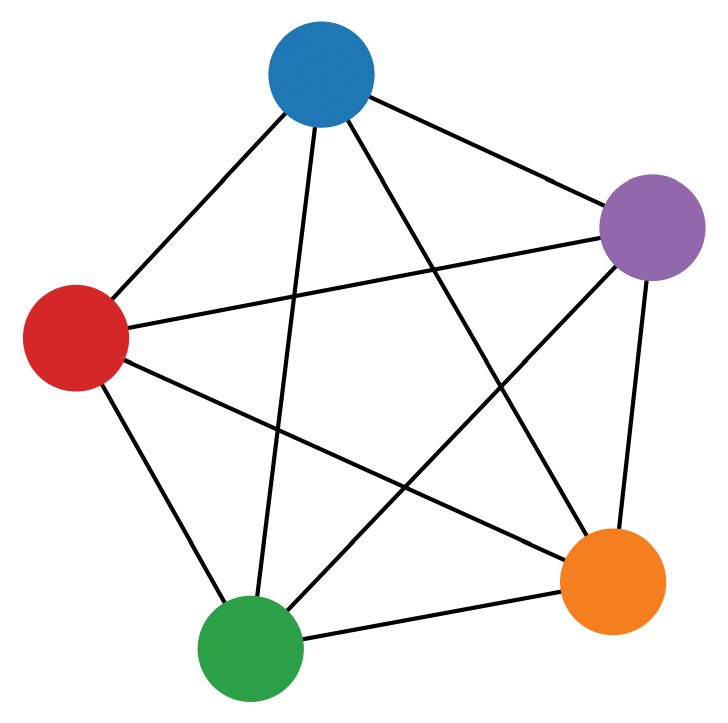}
    \includegraphics[width=0.3\columnwidth]{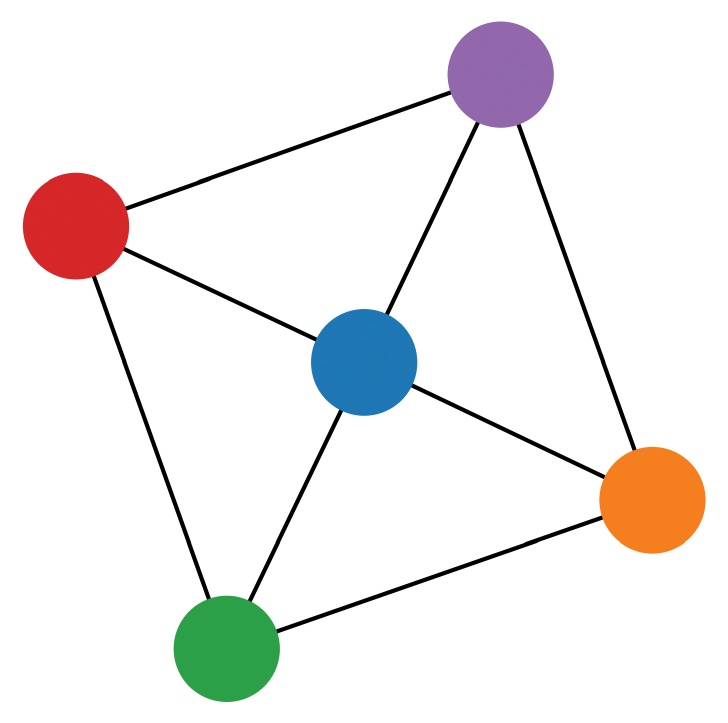}
    \includegraphics[width=0.3\columnwidth]{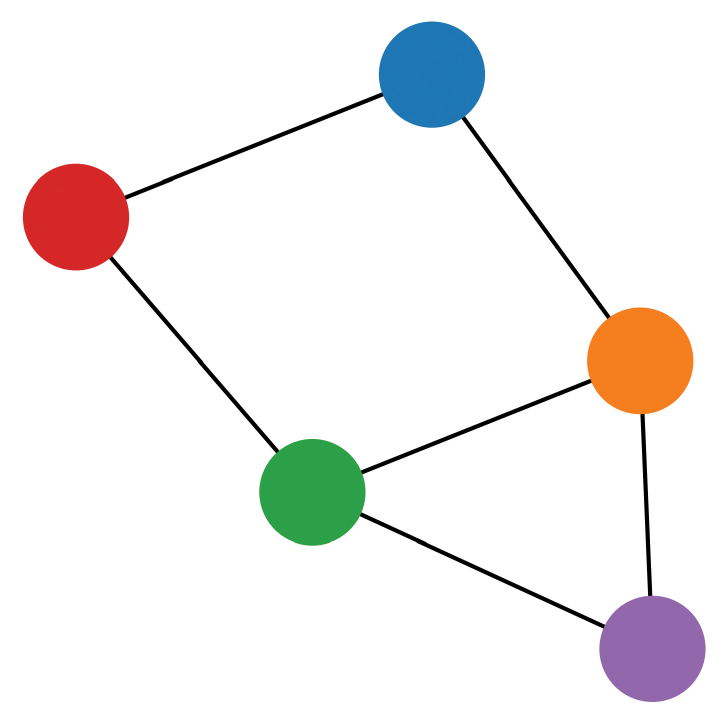}
    \includegraphics[width=0.3\columnwidth]{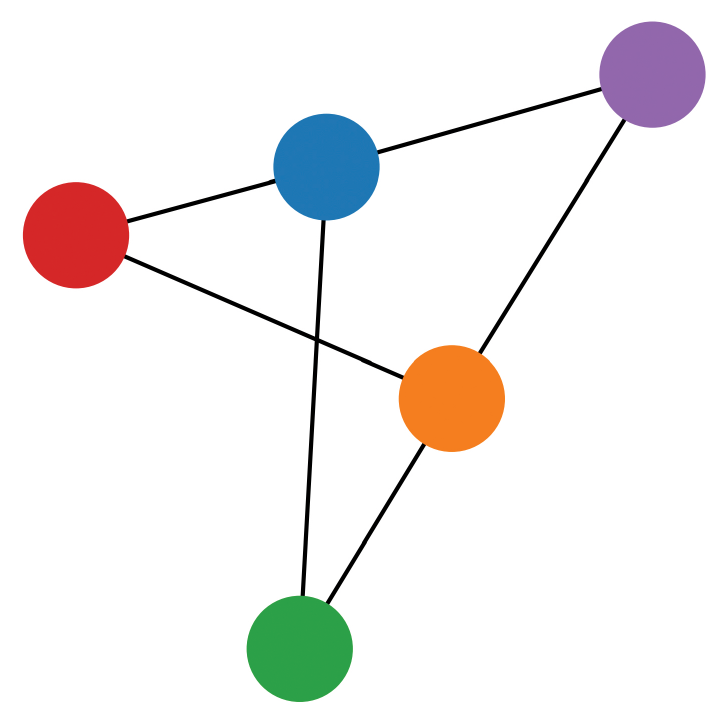}
    \includegraphics[width=0.3\columnwidth]{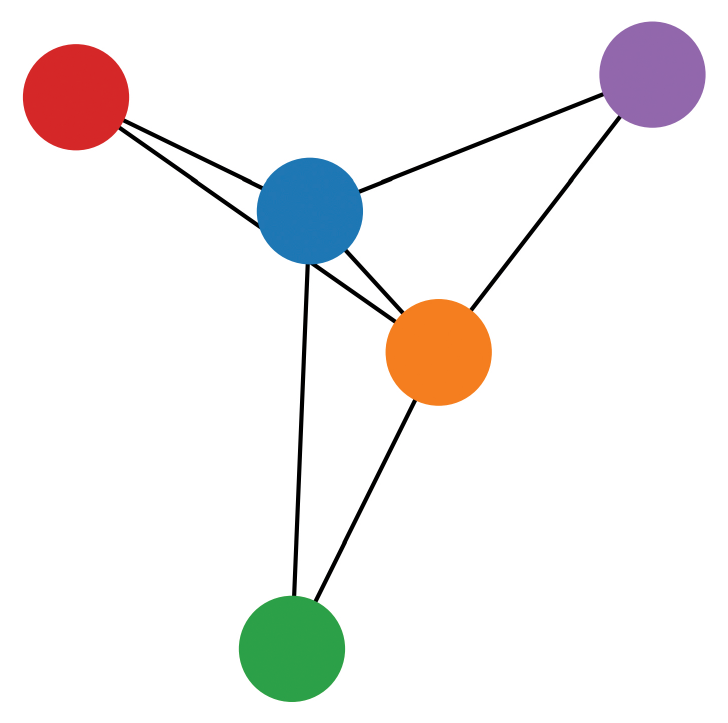}
    \includegraphics[width=0.3\columnwidth]{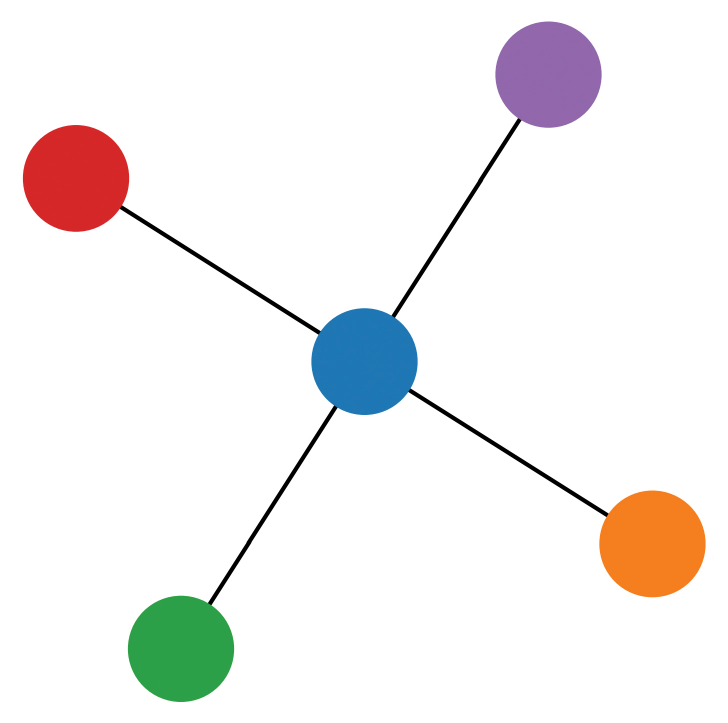}
    \includegraphics[width=0.3\columnwidth]{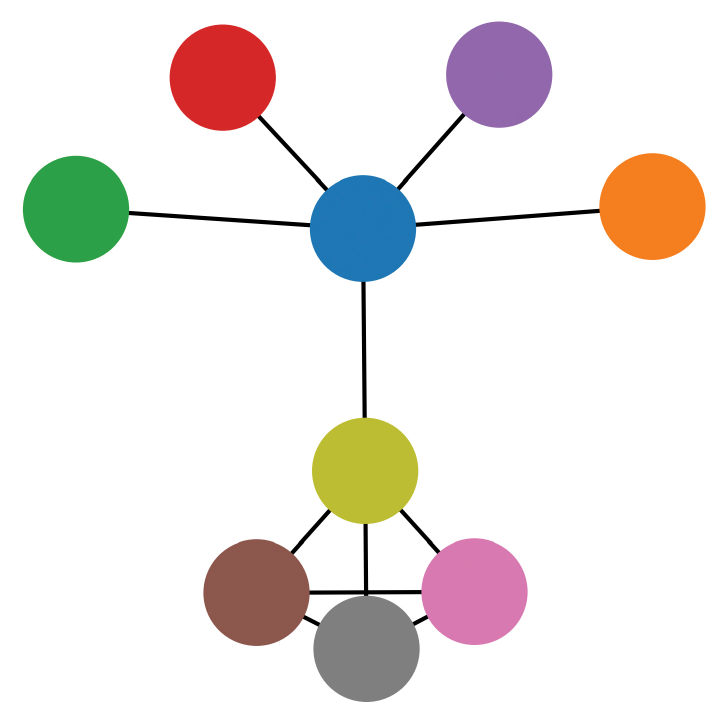}
    \includegraphics[width=0.3\columnwidth]{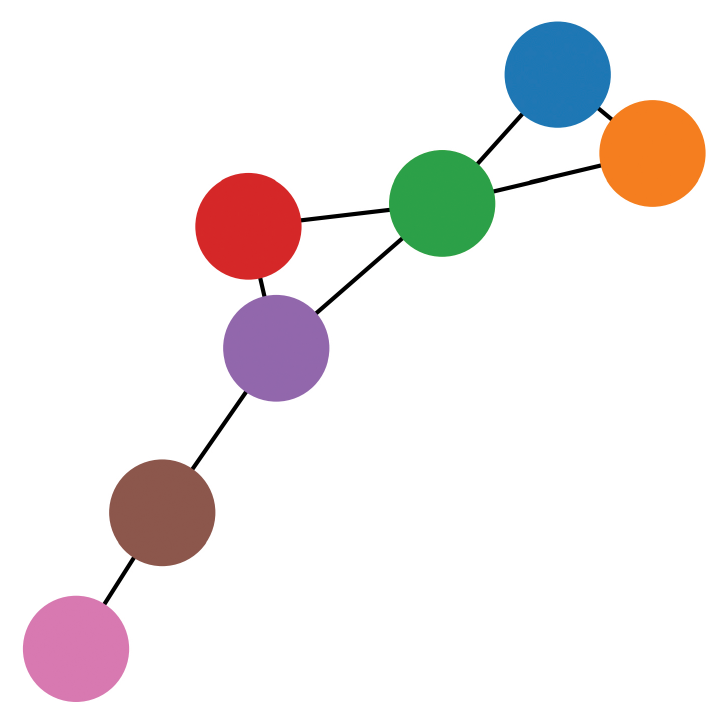}
    \includegraphics[width=0.6\columnwidth]{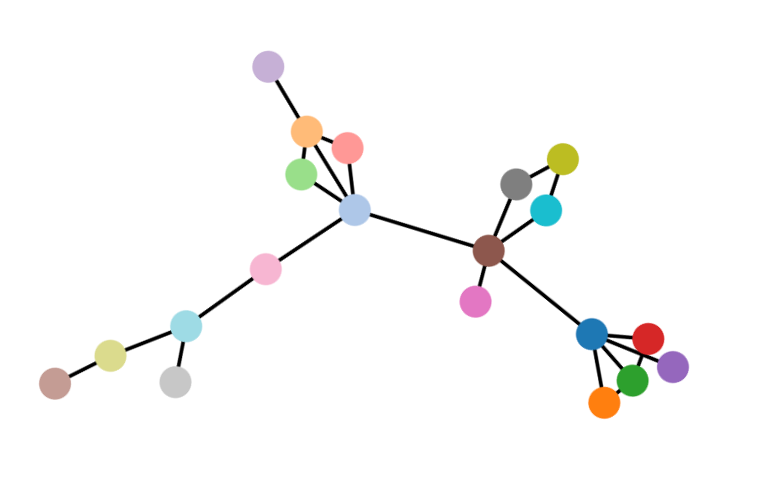}    
    \includegraphics[width=1\textwidth]{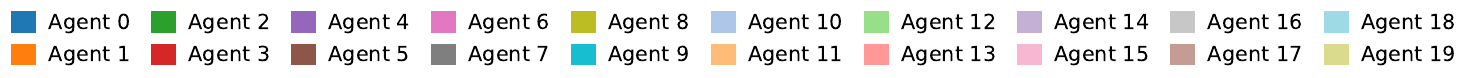}
    \caption{Network Configurations. The first row presents the 5-agent underlying network topologies from left to right:Complete, Wheel, House, Bipartite, Symmetric, and Star. The second row presents the additional three customized network structures: A1, A2, SBM from left to right.}
    \label{fig:topologies}
\end{figure*}

Agents in the Harvest game can take action based on their current partial observation of the state of the environment, where a total of 8 actions can be taken: move forward, backward, left, right, or rotate left or right, fire beam, or keep still. The ``FIRE'' beam action is a punishment action against other agents. When an agent chooses to fire a beam at other agents in one step, in return, it would have one reward unit deducted. Meanwhile, when an agent is hit by a fire beam from another agent, that agent would be fined 50 reward units. This setup is followed by \cite{hughes2018inequity}. The apples in the Harvest's spawn rate is the same setting as in \cite{hughes2018inequity}. 

Similarly, in the Cleanup game, each agent's goal is to collect apples. The apples' respawn rate is controlled by a separate aquifer that accumulates waste over time, linearly reducing the apple respawn rate until spawning stops beyond a saturation threshold. Agent has an additional ``CLEAN'' beam action that removes waste from the aquifer, enabling apples' sustainable regeneration. When the waste level reaches a sufficiently high value, no more apples can spawn. The apples in the Cleanup environment's spawn rate is the same setting as in \cite{hughes2018inequity}.

We expand the sequential social dilemma Markov game with the relational structure to the N-player socio-relational networked partially observable Markov game $\mathcal{M'}$. In the computational implementation of the model, we design each agent's socio-relational structure as an object, class Network, that captures the actual relationship and links between agents under a given underlying topology. Additionally, given a specific network topology defined at the beginning of the game, each agent can access its preferred neighbor portfolio from the Network class. Also, we defined the IDMapping class to ensure the correct identity synchronization between agents' IDs and the agents' node IDs within a topology across all the networked systems. Similar to the implementation of the inequity aversion, the core intrinsic reward is defined and integrated at the environment level through modifying the MapEnv class. With this design, we can implement and examine different homogeneous agents' preferences over sub-graphical structures as environmental modification. 

The experiment model utilizes the default baseline neural network architecture designed in \cite{hughes2018inequity} for agent training and learning, to study the topological structural impact and agent preference impact on MARL agent and system behavior. 
The complete neural network is composed of a convolutional encoder (kernel=3, stride=1, channels=6), and two dense layers (32 neurons), with a Long Short Term Memory (LSTM) actor critic network (128 hidden units). The discount factor is 0.99. The model parameters are identical to the default setup \cite{jaques2019social}. Learning Agents are trained using the Proximal Policy Optimization (PPO) algorithm. The training and data collection are built on the Ray RLlib distributed reinforcement learning framework. Each combination of the underlying topologies and agents' preferences was first evaluated through 5 independent randomized experimental trials until explicit and behavioral patterns stabilized around $3\times 10^7$ agent steps, followed by a full complete experimental trial with standard benchmark duration ($1\times 10^8$) executed repeatedly across 5 random seeds. Within each trial, 20 parallel environments independently operate with random initialization to generate agents' reward and metrics information data for both Harvest and Cleanup. The data collection includes each agent's base reward, the corresponding socio-relational reward in each episode, ``Fire" count as individual aggressiveness, ``Clean" count as individual public good contribution effort,
and the apple counts as the total apple collected in one complete episode. 
Moreover, the data collection also involves system-level metrics that track the collective outcomes and our newly proposed metric of the collective bridging capacity index.

To investigate the impact of socio-relational intrinsic reward, defined based on agents' social preferences over sub-graphical structures, a series of experiments is designed across seven different 5-agent networks and three customized multi-agent network configurations with increasing population size of 7, 9, 20. The analyzed networks are shown in Figure~\ref{fig:topologies} in the supplementary materials. The networks are selected to capture sub-graphical structure variations.

\section{Experimental Result Detail}

\subsection{Structural Evaluation}
\subsubsection{BCI Metric under Nearest Neighbor Preference}

For each topology, the nodes' Burt's constraint value varies. Hence, the bridging capacity of each topology is also distinct at convergence. Specifically, the topological structure inherently determines that there exists a specific lower and upper bound for the bridging capacity of the graph in the Networked Harvest and Cleanup game. Here, the absolute theoretical limit derived from the BCI equation is [0,1]. The bridging potential is defined to be the topology-constrained theoretical range, which represents the full range of feasible BCI values under that specific topological structure. The actual observed range is the values measured from the BCI data collected. We define a topology-constrained theoretical range to be wide if it spans greater than $50\%$ of the absolute theoretical limit, and to be narrow if it spans less than $20\%$ of the absolute theoretical limit. By examining each topology in detail in Harvest, it can be observed that for some topological structures, even though they have a wide range of bridging potential, such as the Star Network with a bridging potential [0,0.75], spanning $75\%$ of the absolute theoretical limit, the actual bridging capacity mean is only 0.023. The actual bridging capacity is $100\%$ lower than 0.100, given that the highest value in bridging potential is 0.750.%

For a network with 5 agents, 7 distinct network topology structures are studied under both Harvest and Cleanup environment. From Table II in the main manuscript, topologies Bipartite, Symmetric, and Star all have their bridging capacity mean and 75th percentile that lie close to the lower bound of the bridging potential. For the Star network, its bridging potential upper bound is 0.75. However, the actual bridging capacity mean is 0.023. This shows that networked agents trained and learned in the Harvest environment with this Star structure collect resources strategically, carrying less from the node with the highest bridging capacity at the bridging position, and more from the node with the lowest bridging capacity, compared to the total resources collected at convergence. Similarly, the structure Bipartite and Symmetric indicates the same pattern of agents' resource collection pattern. For the topology House, the bridging capacity mean is in the middle of the range, which indicates that among the networked trained agents under this topology, the agent with the least bridging capability is able to carry more resources. Meanwhile, agents with high bridging capacity exhibit balanced distributed resources across nodes with different bridging capability, hence the collective BCI value lies in the middle. At the same time, for the topology Wheel, the mean value of the bridging capacity and the 25th percentile lie close to the upper bound of the bridging potential. These indicate that the networked agents with this Wheel structure strategically collect resources more from the agent with high brokerage opportunities who spans the structural hole at the bridging position, as a percentage of the total resources collected at convergence. Hence, brokers with low Burt's constraint create a large portion of the social welfare under the topology Wheel. For the topologies Complete and Cycle, the median, mean, Q1, and Q3 values are exactly the same. This is due to the Complete and Cycle topologies are completely symmetric, and each node has the same bridging capacity. Hence, the collective bridging capacity indexes are exactly the same as the complementary part of Burt's constraint, which is $1-C_{i}$, which is 0.234 for Complete and 0.500 for Cycle.

\subsubsection{BCI Metric under Different Agent Preference With Statistical Tests}
Different topologies with more than 5 agents are also tested. Similar patterns are observed when changing agents' preferences impact agents' learning and lead to distinct strategic behavior under nearest neighbor, Clique-neighbor, and Critical-Connection-neighbor social preferences under A2 configuration. In Figure 2 in the manuscript, A2 with $N=7$ node and A1 with $N=9$ node topology under different preferences are presented, a significant difference in the value of BCI metrics is observed. (Detailed analysis please see the following pairwise comparison). Overall, the agents' preference over the brokerage structure always leads to the brokers gaining the most resources compared to others in the system. While under traditional preference, the agent with the most connections is exploited the most. These distinct BCI metrics across different agent preferences reflect the significance of studying agents' social preferences, even when agents are connected under the same topology and same apple growth rate. And it shows how agents' emergence behavior changes and how the equilibrium shifts with different agents' social preferences.

A pairwise comparison using a t-test with Bonferroni correction ($\alpha = 0.05$) is conducted for the BCI metric under Nearest-Neighbor (NN), Clique-Neighbor (CN), and Critical-Connection neighbor (HBN) preferences for the Star Topology across 5 seeds in Harvest: Nearest-Neighbor preference vs. Clique-Neighbor preference with t$=-22.346$, Bonferroni-adjusted p ($p_{\text{adj}}$) $=0.0000$,  Nearest-Neighbor preference  vs. Critical-Connection neighbor preference $t= -27.9693$, $p_{\text{adj}}$ $=0.000002$, Clique-Neighbor preference vs. Critical-Connection neighbor preference t$=-14.1234$, $p_{\text{adj}}$ $=0.000018$.
All comparisons show statistically significant for multiple testing across different agent preferences ($p_{\text{adj}}<\alpha = 0.05$). With a $95\%$ confidence interval under  Nearest-Neighbor (NN) preference is [0.0149, 0.0345],  $95\%$ confidence interval under Clique-Neighbor (CN) preference is [0.1409, 0.1587], $95\%$ confidence interval under Critical-Connection neighbor (HBN) preference is [0.2631, 0.3114]. Similarly, we compare BCI under Nearest-Neighbor, Clique-Neighbor, and Critical-Connection neighbor preferences for the House, A1, A2 topologies, as well as the SBM topology with 20 agents. All comparisons are significantly different across different agent preferences under A1, A2, SBM topologies. Specifically, for SBM topology, the $95\%$ confidence interval under  Nearest-Neighbor preference is [0.2586, 0.2882], Clique-Neighbor preference is [0.3207, 0.3408], under Critical-Connection neighbor preference is [0.3903, 0.4179] in Harvest. 
For House topology, BCI under Clique-Neighbor preference vs Critical-Connection neighbor preference, Clique-Neighbor preference vs. Nearest-Neighbor preference are both statistically significant. However, no significant difference was detected between  Nearest-Neighbor and Critical-Connection neighbor preference ($0.0643>\alpha$) under Harvest. 


A pairwise comparison using a t-test with Bonferroni correction ($\alpha = 0.05$) is conducted for the BCI metric under Nearest-Neighbor (NN), Clique-Neighbor (CN), and Critical-Connection neighbor (HBN) preferences for the Star Topology across 5 seeds in Cleanup: Nearest-Neighbor preference vs Clique-Neighbor preference with t$=-4.3707$, Bonferroni-adjusted p ($p_{\text{adj}}$) $=0.0359$, Nearest-Neighbor preference vs Critical-Connection neighbor preference $t=-8.7598$, $p_{\text{adj}}$ $=0.0028$, Clique-Neighbor preference vs. Critical-Connection neighbor preference t$=-4.3384$, $p_{\text{adj}}$ $=0.0088$.
All comparisons show statistically significant for multiple testing across different agent preferences ($p_{\text{adj}}<\alpha = 0.05$). With a $95\%$ confidence interval under Nearest-Neighbor preference is [0, 0], Clique-Neighbor preference is [0.0683, 0.3060], Critical-Connection neighbor preference is [0.3378, 0.6513]. Similarly, we compare BCI under Nearest-Neighbor, Clique-Neighbor, and Critical-Connection neighbor preferences for the House, A1, A2 topologies. All comparisons are significantly different across different agent preferences under A1, A2. 
For House topology, BCI under Nearest-Neighbor preference vs Critical-Connection neighbor preference, Clique-Neighbor preference vs. Nearest-Neighbor preference are both statistically significant. However, no significant difference was detected between  Clique-Neighbor preference vs Critical-Connection neighbor preference ($p_{\text{adj}}=1>\alpha$) under House topology in Cleanup.

\subsubsection{Pairwise comparison under Harvest}

\paragraph {Star Topology} A pairwise comparison using the t-test with Bonferroni correction ($\alpha = 0.05$) is conducted for the BCI metric under Nearest-Neighbor (NN), Clique-Neighbor (CN) , and Critical-Connection neighbor (HBN) preferences for the Star Topology in Harvest environment across 5 seeds: Nearest-Neighbor (NN) vs. Clique-Neighbor (CN) preference  with $t=-22.3460$, $p_{\text{adj}}$$=0.0000$, Nearest-Neighbor (NN) preference vs. Critical-Connection neighbor (HBN) preference $t=-27.9693$, $p_{\text{adj}}$$=0.0.000002$, Clique-Neighbor (CN) preference vs. Critical-Connection neighbor (HBN) preference $t=-14.1234$, $p_{\text{adj}}$$=0.000018$. By comparing the Bonferroni-corrected p-value comparison with the significance threshold $\alpha = 0.05$, all comparisons show significance after correction for multiple testing across different agent preferences. With a $95\%$ confidence interval under  Nearest-Neighbor (NN) preference is [0.0149, 0.0345],  $95\%$ confidence interval under Clique-Neighbor (CN) preference is [0.1409, 0.1587], $95\%$ confidence interval under Critical-Connection neighbor (HBN) preference is [0.2631, 0.3114].

\paragraph {House Topology} A pairwise comparison using the t-test with Bonferroni correction ($\alpha = 0.05$) is conducted for the BCI metric under Nearest-Neighbor (NN), Clique-Neighbor (CN), and Critical-Connection neighbor (HBN) preferences for the House Topology across 5 seeds: Nearest-Neighbor (NN) preference vs. Clique-Neighbor (CN) preference with t=$-5.6188$, $p_{\text{adj}}$$=0.0019$, Nearest-Neighbor (NN) preference vs. Critical-Connection neighbor (HBN) preference t$=-2.7232$, $p_{\text{adj}}$$=0.0643$, Clique-Neighbor (CN) preference vs. Critical-Connection neighbor (HBN) preference t$=3.8738$, $p_{\text{adj}}$$=0.0163$. From pairwise comparison,BCI under Clique-Neighbor (CN) and Critical-Connection neighbor (HBN) are significantly different, BCI under Clique-Neighbor (CN) vs Nearest-Neighbor (NN) are significantly different. However, no significant difference was detected between  Nearest-Neighbor (NN) and Critical-Connection neighbor (HBN) preferences $(0.0643>\alpha = 0.05)$ With a $95\%$ confidence interval under  Nearest-Neighbor (NN) preference is [0.3620, 0.3863], Clique-Neighbor (CN) preference is [0.4084, 0.4551], Critical-Connection neighbor (HBN) preference is [0.3802, 0.4041].

\paragraph {A2 Topology $N=7$} A pairwise comparison using the t-test with Bonferroni correction ($\alpha = 0.05$) is conducted for the BCI metric under Nearest-Neighbor (NN), Clique-Neighbor (CN) , and Critical-Connection neighbor (HBN) preferences for the A2 Topology across 5 seeds:  Nearest-Neighbor (NN) preference vs. Clique-Neighbor (CN) preference with t$=-3.8753$, $p_{\text{adj}}$$=0.0161$, Nearest-Neighbor (NN) preference vs. Critical-Connection neighbor (HBN) preference  t$=-7.9488$, $p_{\text{adj}}$$=0.00004$, Clique-Neighbor (CN) preference vs. Critical-Connection neighbor (HBN) preference t$=-6.1162$, $p_{\text{adj}}$$=0.00103$. By comparing the Bonferroni-corrected p-value comparison with the significance threshold $\alpha = 0.05$, all comparisons show significance after correction for multiple testing across different agent preferences. With a $95\%$ confidence interval under  Nearest-Neighbor (NN) preference is [0.135281, 0.196928], Clique-Neighbor (CN) preference is [0.202504, 0.234216], Critical-Connection neighbor (HBN) preference is [0.268797, 0.328254].

\paragraph {A1 Topology $N=9$} A pairwise comparison using the t-test with Bonferroni correction ($\alpha = 0.05$) is conducted for the BCI metric under Nearest-Neighbor (NN), Clique-Neighbor (CN), and Critical-Connection neighbor (HBN) preferences for the A1 Topology across 5 seeds: Nearest-Neighbor (NN) preference vs, Clique-Neighbor (CN) preference with t$=-11.2555$, $p_{\text{adj}}=0.000022$,  Nearest-Neighbor (NN) preference vs. Critical-Connection neighbor (HBN) preference t$= -23.8851$, $p_{\text{adj}}=0.0000$, Clique-Neighbor (CN) preference vs. Critical-Connection neighbor (HBN) preference t$= -15.4519$, $p_{\text{adj}}=0.000001$. By comparing the Bonferroni-corrected p-value comparison with the significance threshold $\alpha = 0.05$, all comparisons show significance after correction for multiple testing across different agent preferences. Specifically, for A1 topology, the $95\%$ confidence interval under  Nearest-Neighbor preference is [0.0379, 0.0821], Clique-Neighbor preference is [0.1546, 0.1866], under Critical-Connection neighbor preference is [0.2847, 0.3195].

\paragraph {SBM Topology with $N=20$} A pairwise comparison using the t-test with Bonferroni correction ($\alpha = 0.05$) is conducted for the BCI metric under Nearest-Neighbor (NN), Clique-Neighbor (CN), and Critical-Connection neighbor (HBN) preferences for the SBM Topology across 5 seeds: Nearest-Neighbor (NN) preference vs, Clique-Neighbor (CN) preference with t$=-12.3066$, $p_{\text{adj}}=0.000005$,  Nearest-Neighbor (NN) preference vs. Critical-Connection neighbor (HBN) preference t$= -22.2424$, $p_{\text{adj}}=0.0000$, Clique-Neighbor (CN) preference vs. Critical-Connection neighbor (HBN) preference t$=-11.9538$, $p_{\text{adj}}=0.000014$. By comparing the Bonferroni-corrected p-value comparison with the significance threshold $\alpha = 0.05$, all comparisons show significance after correction for multiple testing across different agent preferences. Specifically, for SBM topology, the $95\%$ confidence interval under  Nearest-Neighbor preference is [0.2586, 0.2783], Clique-Neighbor preference is [0.3207, 0.3408], under Critical-Connection neighbor preference is [0.3903, 0.4179].

\subsubsection{Pairwise comparison under Cleanup}

\paragraph {Star Topology}
A pairwise comparison using the t-test with Bonferroni correction ($\alpha = 0.05$) is conducted for the BCI metric under Nearest-Neighbor (NN), Clique-Neighbor (CN), and Critical-Connection neighbor (HBN) preferences for the Star Topology across 5 seeds: Nearest-Neighbor (NN) preference vs, Clique-Neighbor (CN) preference with t$=-4.3707$, $p_{\text{adj}}=0.0359$,  Nearest-Neighbor (NN) preference vs. Critical-Connection neighbor (HBN) preference t$=-8.7598$, $p_{\text{adj}}= 0.0028$, Clique-Neighbor (CN) preference vs. Critical-Connection neighbor (HBN) preference t$-4.3384$, $p_{\text{adj}}= 0.0088$. By comparing the Bonferroni-corrected p-value comparison with the significance threshold $\alpha = 0.05$, all comparisons show significance after correction for multiple testing across different agent preferences. For Star topology, the $95\%$ confidence interval under  Nearest-Neighbor preference is [0, 0], Clique-Neighbor preference is [0.0683, 0.3060], under Critical-Connection neighbor preference is [0.3378, 0.6513].

\paragraph {House Topology}
A pairwise comparison using the t-test with Bonferroni correction ($\alpha = 0.05$) is conducted for the BCI metric under Nearest-Neighbor (NN), Clique-Neighbor (CN), and Critical-Connection neighbor (HBN) preferences for the House Topology across 5 seeds: Nearest-Neighbor (NN) preference vs, Clique-Neighbor (CN) preference with t$=-3.9128$, $p_{\text{adj}}=0.0368$,  Nearest-Neighbor (NN) preference vs. Critical-Connection neighbor (HBN) preference t$=-3.6497$, $p_{\text{adj}}= 0.0415$, Clique-Neighbor (CN) preference vs. Critical-Connection neighbor (HBN) preference t$=0.7283$, $p_{\text{adj}}=1$. For House topology, BCI under Nearest-Neighbor preference vs Critical-Connection neighbor preference, Clique-Neighbor preference vs. Nearest-Neighbor preference are both statistically significant. However, no significant difference was detected between  Clique-Neighbor preference vs Critical-Connection neighbor preference ($p_{\text{adj}}=1>\alpha$) under Cleanup. Specifically, for House topology, the $95\%$ confidence interval under  Nearest-Neighbor preference is [0.3095, 0.3551], Clique-Neighbor preference is [0.3668, 0.5115], under Critical-Connection neighbor preference is [0.3563, 0.4732].

\paragraph {A2 Topology $N=7$}
A pairwise comparison using the t-test with Bonferroni correction ($\alpha = 0.05$) is conducted for the BCI metric under Nearest-Neighbor (NN), Clique-Neighbor (CN), and Critical-Connection neighbor (HBN) preferences for the A2 Topology across 5 seeds: Nearest-Neighbor (NN) preference vs, Clique-Neighbor (CN) preference with t$= -4.5711$, $p_{\text{adj}}=0.0060$,  Nearest-Neighbor (NN) preference vs. Critical-Connection neighbor (HBN) preference t$= -11.5213$, $p_{\text{adj}}=0.000009$, Clique-Neighbor (CN) preference vs. Critical-Connection neighbor (HBN) preference t$= -5.6042$, $p_{\text{adj}}=0.0019$. By comparing the Bonferroni-corrected p-value comparison with the significance threshold $\alpha = 0.05$, all comparisons show significance after correction for multiple testing across different agent preferences. Specifically, for A2 topology, the $95\%$ confidence interval under  Nearest-Neighbor preference is [0.0778, 0.1655], Clique-Neighbor preference is [0.1821, 0.2889], under Critical-Connection neighbor preference is [0.3305, 0.4132].

\paragraph {A1 Topology $N=9$}
A pairwise comparison using the t-test with Bonferroni correction ($\alpha = 0.05$) is conducted for the BCI metric under Nearest-Neighbor (NN), Clique-Neighbor (CN), and Critical-Connection neighbor (HBN) preferences for the A1 Topology across 5 seeds: Nearest-Neighbor (NN) preference vs, Clique-Neighbor (CN) preference with t$=-14.6451,$, $p_{\text{adj}}=0.0003$,  Nearest-Neighbor (NN) preference vs. Critical-Connection neighbor (HBN) preference t$= -15.5988$, $p_{\text{adj}}= 0.0003$, Clique-Neighbor (CN) preference vs. Critical-Connection neighbor (HBN) preference t$=-6.5251$, $p_{\text{adj}}=0.0016$. By comparing the Bonferroni-corrected p-value comparison with the significance threshold $\alpha = 0.05$, all comparisons show significance after correction for multiple testing across different agent preferences. Specifically, for SBM topology, the $95\%$ confidence interval under  Nearest-Neighbor preference is [0.0332, 0.0407], Clique-Neighbor preference is [0.1289, 0.1711], under Critical-Connection neighbor preference is [0.2143, 0.2907].

 \paragraph {SBM Topology with $N=20$} 
A pairwise comparison using the t-test with Bonferroni correction ($\alpha = 0.05$) is conducted for the BCI metric under Nearest-Neighbor (NN), Clique-Neighbor (CN), and Critical-Connection neighbor (HBN) preferences for the SBM Topology across 5 seeds under Cleanup: Nearest-Neighbor (NN) preference vs, Clique-Neighbor (CN) preference with t$=--30.4407$, $p_{\text{adj}}=0.002405$,  Nearest-Neighbor (NN) preference vs. Critical-Connection neighbor (HBN) preference t$= -17.7444$, $p_{\text{adj}}= 0.003509$, Clique-Neighbor (CN) preference vs. Critical-Connection neighbor (HBN) preference t$=-8.2023$, $p_{\text{adj}}=0.042821$. By comparing the Bonferroni-corrected p-value comparison with the significance threshold $\alpha = 0.05$, all comparisons show significance after correction for multiple testing across different agent preferences. Specifically, for SBM topology, the $95\%$ confidence interval under Nearest-Neighbor preference is [0.1978,0.2255], Clique-Neighbor preference is [0.3086,0.3132], under Critical-Connection neighbor preference is [0.3477, 0.4288].

\subsection{Individual Evaluation}

\subsubsection{Base Reward}

\begin{figure*}[t]
    \centering

    \includegraphics[width=0.30\textwidth]{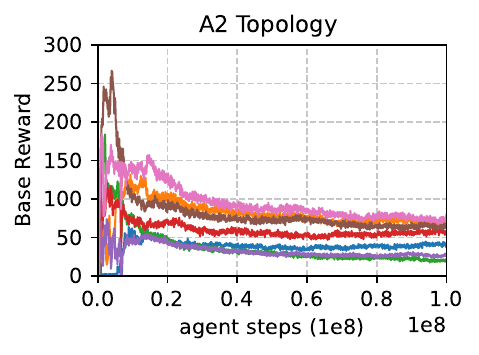}
    \includegraphics[width=0.30\textwidth]{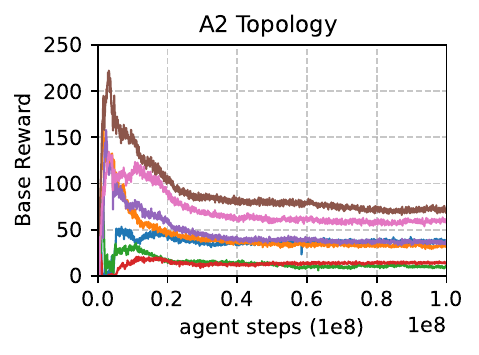}
    \includegraphics[width=0.30\textwidth]{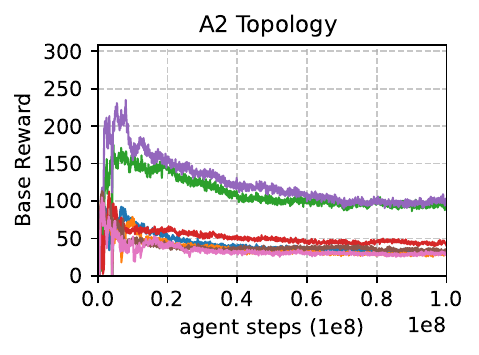}    \includegraphics[width=0.30\textwidth]{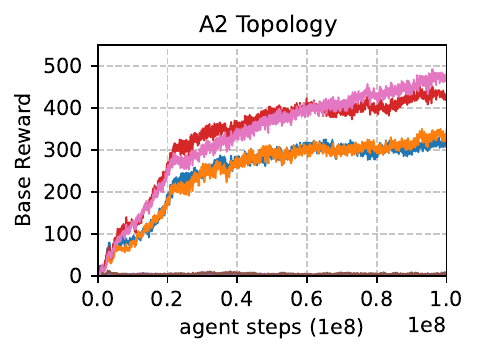}
    \includegraphics[width=0.30\textwidth]{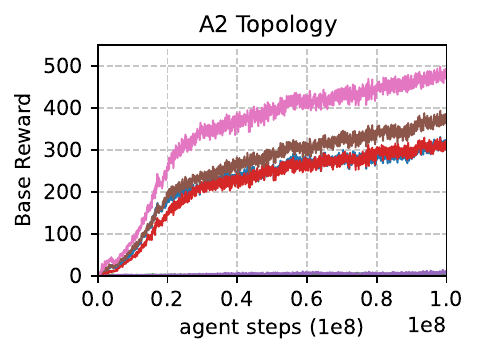}
    \includegraphics[width=0.30\textwidth]{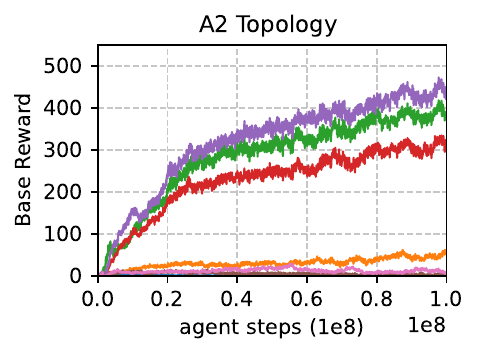}
\caption{\label{fig:basereward_indiv2} Individual mean base reward under different agent preferences under A2 topology with N=7 (Agent 0 to Agent 7). Top Row: Harvest, Bottem Row: Cleanup. $1^{st}$ column: under the Nearest Neighbor Preference ($\alpha=1$); $2^{nd}$  column: Clique-Neighbor Preference ($\beta=1$); $3^{rd}$  column: Critical-Connection Neighbor Preference ($\omega=1$). Non-active parameters set to 0. }
\end{figure*}

Under the Nearest Neighbor Preference, the pattern observed in both Harvest and Cleanup, where the agent that cares for its neighbors the most and the most prosocial agent, with the highest degree centrality, at least one of them obtains the lowest resources compared to others. The pattern is robust across most examined topologies, including A1 and A2, as well as 5-agent network configurations of House, Bipartite, Symmetric, and Star across all 5 seeds. The findings are consistent across degree variations and population rate variations. For topologies like Bipartite and Symmetric, in which two agents hold the highest degree centrality and connect to all agents without the highest degree centrality, both agents obtain the lowest resources across 5 seeds. For topologies like House, in which two agents hold the highest degree centrality while not connecting to all agents without the highest degree centrality, we observed alternating reward trajectories with Agent 1 or Agent 2 alternating to obtain the lowest resources, while the other agent collects the least to moderate resources across 5 seeds.

Figure~\ref{fig:basereward_indiv2} shows the individual base reward trajectory for topology A2, which exhibits consistent recurring trajectories patterns across preferences as Star, House, A1 as shown in manuscript Fig.3 and discussed in Sec.B.1.

Meanwhile, the exception arises for the Wheel topology with peripheral agents caring 3 other neighbors, and hub agents caring all 4 remaining agents under the nearest neighbor preference. In Wheel topologies, all agents care about the majority of other agents and are highly coupled in rewards. With 3 trials, in which the hub agent (Agent 0) collected the least reward, the general findings and patterns were preserved as stated in the main manuscript. And the other 2 trials with Agent 0 collecting the second least or the middle-ranked rewards, and other agents exhibit randomly ordered reward trajectories that are similar to the complete topology's reward trajectories. This indicates that when all agents care for the majority of their neighbors and act mostly prosocially, the care-about neighborhood substantially overlaps among agents, leading to higher variability and randomness in reward collection occurring in the wheel topology (similar to complete topology), along with lower observable sub-graphical structural differentiation, under the nearest-neighbor preference.

\subsubsection{Individual Aggressiveness}

\begin{figure*}[t]
    \centering

    \includegraphics[width=0.8\textwidth]{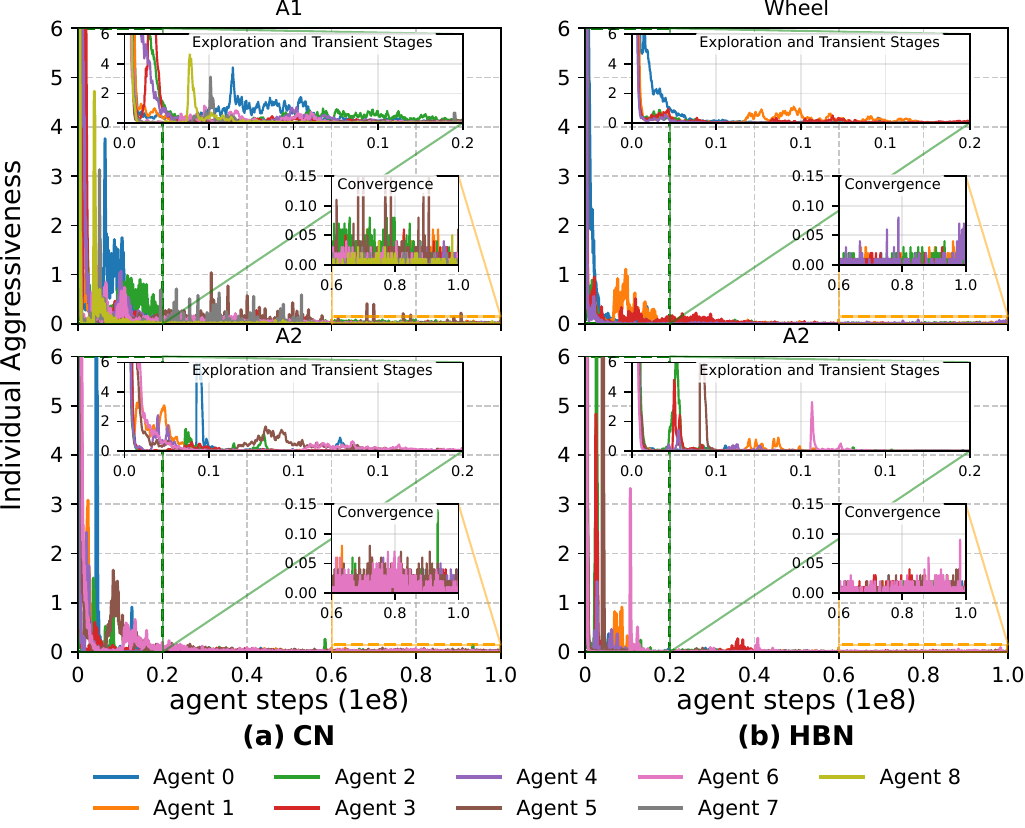}
    
    \caption{This figure shows overall agents' individual aggressiveness over the entire experimental agent steps under (a) clique neighbor preference (CN) in 9-agent A1 and 7-agent A2 topology, and (b) critical-connection preference (HBN) in 5-agent Wheel and 7-agent A2 topology. The upper subregion subfigure shows agents' individual aggressiveness in random exploration and transient stages with x-axis limit at $2e7$ and original y-scale. This lower subregion subfigure shows agents' individual aggressiveness at convergence, with the x-axis range: $[0.6e8,1e8]$, and y-axis range: $[0,0.15]$.}
    \label{fig:inv_fire_count_CN_highlight2}

\end{figure*}
\paragraph{Aggressiveness under Traditional Neighbor Preference}
 In each 5-agent network configuration, at least one agent with the highest degree centrality kept its mean summed aggressiveness at the highest or second-highest level among all agents and consistently ranked within the top-2 highest aggressiveness in every seed, except the Star topology. In Star Topology, in two seeds where the agent (Agent 0) with the highest degree centrality successfully learn the complicated cooperative policy of sanction, the mean summed aggressiveness over the two seed is 37.02, which is the highest mean summed aggressiveness among all agent, with other agents' aggressiveness from agent 1 to agent 4 are 12.18, 28.28, 7.39, 6.11 accordingly. Meanwhile, for the other 3 seeds, where Agent 0 exhibits failure to learn the sanction policy, with the mean summed aggressiveness over the three seeds are 6.18, which is the lowest mean summed aggressiveness among all agents, with other agent's aggressiveness from agent 1 to agent 4 are 20.98, 22.01, 31.86, 18.14 accordingly. The mean summed aggressiveness across seeds shows that agents on the periphery exhibit elevated aggressiveness and act selfishly when Agent 0 fails to learn the sanction policy.

\paragraph{Aggressiveness under Clique Neighbor Preference}

Figure~\ref{fig:inv_fire_count_CN_highlight2}(a) in this supplementary materials shows the results that the agents who form clique alliances and care about the clique structure preference  (Agent 5-8 in A1 topology; Agent 0-4 in A2 topology) learn to reduce their aggressiveness the fastest in the first exploration stage, in contrast to agents outside the clique alliance (Agent 0-4 in A1 topology and Agent 5, 6 in A2 topology).  As learning proceeds, some agents in clique alliances (Agent 5, 7, 8 in A1 and Agent 0, 1, 2 in A2 topology) undergo a transient stage characterized by a sudden increase in the frequency of applying the ``FIRE'' action, followed by readjusting their aggressiveness to a lower level. This transient behavior indicates that those agents strategically alter their non-aggressive prosocial behavioral policy learned in the first stage and explore the increased utilization of aggressiveness, while striving to maximize their objectives of the clique alliance benefit. At convergence, within the clique alliance, some agent (Agent 5, 6 in A1, and Agent 1, 2 in A2 topology) consistently exhibited aggressiveness typically $3\times$ higher than that of other clique alliance agents in A1 and A2 with rare cases ranging from $2\times$ to $5\times$ for House, A1 and A2 across 5 seeds.  In House Topology, the agent maintained higher aggressiveness with mean summed aggressiveness is 20.11, whereas the other alliance agents with mean summed aggressiveness is 4.72 across seeds. In A2, the agent maintained higher aggressiveness with mean summed aggressiveness is 14.25, whereas the other alliance agents with mean summed aggressiveness is 4.17. In A1, the agent maintained higher aggressiveness with mean summed aggressiveness is 19.92, whereas the other alliance agents with mean summed aggressiveness is 4.78. This higher aggressiveness is compatible with their clique-alliance group objectives and also costly to their own benefit, which indicates this aggressiveness as an altruistic punishment policy. In contrast, most agents outside the clique alliances (Agent 1, 2, 3 in A1 and Agent 5, 6 in A2) exhibit a consistently higher level of aggressiveness compared to the agents who formed the clique alliances over the entire experimental period. The mean summed aggressiveness of non-clique alliance agents is 20.25, 23.17, and 22.25 for House, A2, and A1 topologies across 5 seeds. This finding aligns with the base reward trajectory, where agents outside the clique structure gather more resources due to their selfish exploitation and their higher level of competition over resources.

\paragraph{Aggressiveness under Critical-Connection Neighbor Preference}

For the scenario of agents with preference over the critical connection path neighbor under the Wheel structure, a pattern can also be observed where the node agent 0 with the highest betweenness value the Wheel topologies learned to reduce aggressiveness relatively slower compared to the agents on the periphery during the early exploration trials, which is shown in the upper subregion of the upper subfigure of Figure \ref{fig:inv_fire_count_CN_highlight2}(b) in supplementary materials. Complicated topology like A2 with 7-agent, which is shown in the bottom subfigure of Figure \ref{fig:inv_fire_count_CN_highlight2}(b) in supplementary materials, with agent 4 strategically positioned as brokers, exhibits the least aggressiveness throughout the entire experimental steps. While Agent 2 in A2 is strategically positioned as a hub node, it exhibits similar patterns of low aggressiveness at the convergence stage. Combined with the individual base reward, Agent 4 and Agent 2, who function as critical connectors, collect the most resources but maintain low aggressiveness at convergence. A similar key insight can be derived and align with the analysis in the main manuscript. The mean summed aggressiveness of agents that function as critical connectors is 1.6 for Agent 0 in Wheel, 3.43 for Agent 0 in Star, and 2.13 for Agent 2 in A2 and 1.62 for Agent 4 in A2 across 5 seeds at convergence, which are the least to second-least mean summed aggressiveness compared to other agents in the same topology. Meanwhile, for A1 with N=9, the mean summed aggressiveness for Agent 0 and Agent 8 is 12.97 and 13.09, respectively, and the mean summed aggressiveness for agents not functioning as critical connectors is 11.22, with a range of [4.49, 25.62]. Hence, the agents functioning as critical connectors exhibit a moderate level of aggressiveness across 5 seeds. The investigation of individual aggressiveness under the same topology demonstrated that changes in agents’ preferences over the microscopic structure produce observable strategic behavioral variations.

\subsubsection{Individual Contribution Effort}

\begin{figure*}[t]
    \centering

    \includegraphics[width=0.30\textwidth]{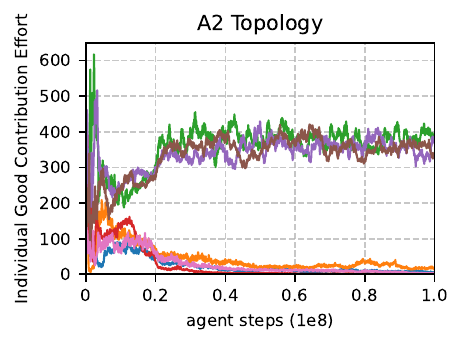}
    \includegraphics[width=0.30\textwidth]{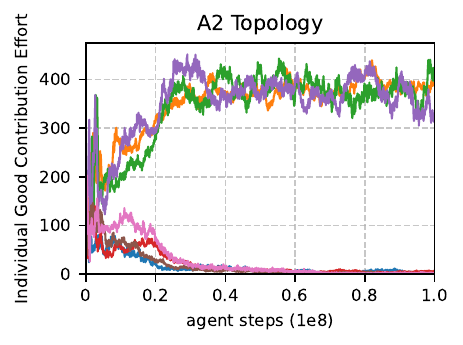}
    \includegraphics[width=0.30\textwidth]{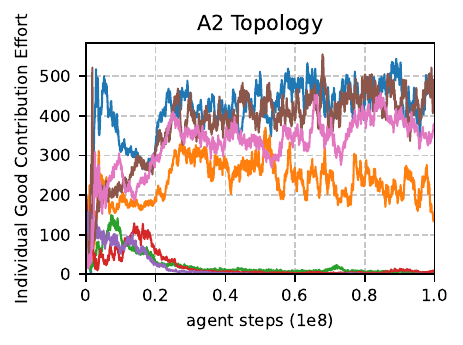}

\caption{\label{fig:Clean_Count2} Individual Good Contribution under different agent preferences under A2 in Cleanup. $1^{st}$ column: Nearest Neighbor Preference; $2^{nd}$ column: Clique-Neighbor Preference; $3^{rd}$ column: Critical-Connection Neighbor Preference. Colors are consistent with Fig.3 legend in manuscript. }
\end{figure*}

Figure~\ref{fig:Clean_Count2} shows the individual good contribution effort under topology A2, which exhibits consistent recurring cleaning trajectories patterns across preferences as Star, House, A1 as shown in manuscript Fig.5 and discussed in Sec.B.3.

\paragraph{SCI under Clique Neighbor Preference} 
 In House, A1 and A2 topologies, all the agents with the highest individual contribution effort are agents who form clique alliances, with $95\%$ confidence interval of SCI for cleaner role agents is $[0.998,1]$ with mean SCI is $0.999$ under House topology, $[0.963,1]$ with mean SCI is $0.983$ under A1, and $[0.996,1]$ with mean SCI is $0.998$ under A2. Meanwhile, agents serve as harvesters in the clique structure with a $95\%$ confidence interval of SCI is $[-0.004,0.01]$ with a mean SCI is $0.0069$ under House topology, $[-0.069,0.402]$ with a mean SCI $0.166$ under A1, and $[0,0.048]$ with a mean SCI $0.024$ under A2. The Star topologies have a $95\%$ confidence interval of SCI is $[0.54,0.79]$ for all agents with a mean SCI $0.67$, since the agents have no clique structure, agents randomly adopt the cleaner role across 5 seeds with no consistent pattern and role specialization observed.

 \paragraph{SCI under Critical-Connection Neighbor Preference} 
 The mean SCI of the highest betweenness agents in S is $0.0075$ with a $95\%$ confidence interval of SCI is $[0.005, 0,009]$. The mean SCI of H, A1, A2 topologies are $0.0112$, $0.0207$ and $0.0322$, with a $95\%$ confidence interval of SCI is $[0.006,0,016]$, $[0.0009,0.041]$ and $[0.013,0.052]$ accordingly. Those SCI confidence intervals close to 0 reflect that agents with the highest betweenness centrality acted purely selfishly and consistently served a harvester role across all timesteps after convergence, across all topologies, across all seeds. This shows that the individual trade-off between its prosocial behavior and self-interested behavior is consistent and recurring across all topologies under each subgraphical preference. Meanwhile, under the same topology, changes in agents' preferences over the subgraphical structure lead to observable consistent variations in agents' strategic behavior.

\subsection{Model Robustness}
\subsubsection{Model Robustness: Dynamic Weight Scheduling}

\begin{figure}[t]
    \centering
    \includegraphics[width=0.78\columnwidth]{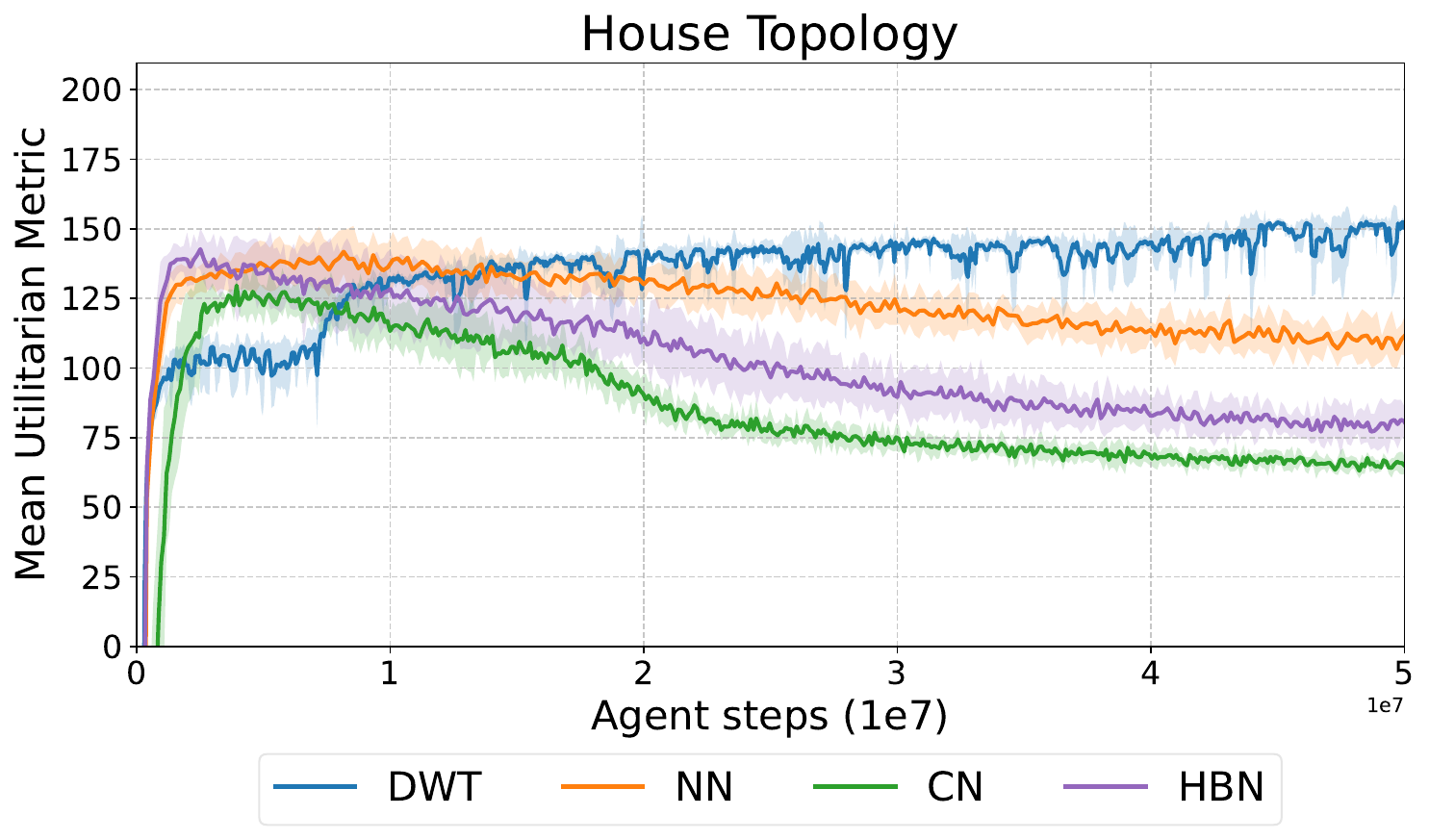}
    \caption{\label{fig:H_U_Compare} Mean utilitarian metric under House topology with different agent preferences in Harvest. 'DWT' represents the dynamic weights tuning over mixed agent preferences by a population-based training approach, `NN': nearest-neighbor preferences, `CN': the clique-neighbor preferences, `HBN': the critical-connection neighbor preferences. Error Bars illustrate a $95\%$ confidence intervals over 5 random seeds. }

\end{figure}

A dynamically changing mixed preference was implemented using the Population-Based Training algorithm to simulate a complex mixed preference optimization with the same population and topological constraints in the Harvest environment, demonstrating our model as a foundation for future research on dynamic weights and further optimization over collective performance. Figure \ref{fig:H_U_Compare} demonstrates the convergence of utilitarian metrics across different agent preferences over sub-graphical structures in House Topology with 5 random seeds. We utilize Population-Based Training as the dynamic hyperparameter optimization method to simultaneously tune the $\alpha, \beta, \omega$ weights across different agent preferences. The dynamic weight tuning approach achieved a mean utilitarian metric of $150.549$ with a standard deviation of $3.799$ and a $95\%$ confidence interval of [148.771, 152.327] across 20 random seeds. A pairwise comparison using the t-test with Bonferroni correction ($\alpha = 0.05$) is conducted for the utilitarian metric under dynamic weights tuning (DWT), Nearest-Neighbor (NN), Clique-Neighbor (CN), and Critical-Connection neighbor (HBN) preferences for the House Topology across 5 seeds. Pair-wise comparisons reveal that the dynamic weights tuning approach significantly outperforms all individually activated sub-graphical preferences from a collective performance (social welfare) perspective. This suggests that collective return optimization is plausible with dynamic scheduling or a mixed combination of weights over agent preferences. Consequently, the selection of teams of agents with distinct sub-graphical preferences may lead to optimal collective performance, even though optimizing social welfare is not the primary objective of this study. All comparisons pairs show statistically significant for multiple testing across different agent preferences($p_{\text{adj}}<\alpha$). With a $95\%$ confidence interval under dynamic weights tuning (DWT) preference is [148.7708, 152.3272]. Meanwhile, pairwise comparisons with all $p_{\text{adj}}<0.0003<\alpha$ under Nearest-Neighbor (NN), Clique-Neighbor (CN), and Critical-Connection neighbor (HBN) indicate that $\alpha$-only, $\beta$-only, $\omega$-only agent preferences differ significantly on the utilitarian metric. The $95\%$ confidence interval under Nearest-Neighbor preference is [104.2768, 120.3084], under Clique-Neighbor preference is [62.0570, 67.6366] and under Critical-Connection neighbor preference is [75.5390, 86.2002]. This shows that the three preferences do lead to qualitatively distinct sub-graphical driven collective performance that warrants separate examination under the Harvest game, as well as other networked SSDs. Furthermore, individually activated sub-graphical preferences are necessary to better disentangle the microstructural impact and identify distinct sub-graphical-driven behaviors, providing sound interpretability for complex situations where agents hold mixed preferences across different sub-graphical structures.

\subsubsection{Model Robustness: Population Rate} 
The investigation of individual behavioral shifts under different topologies across different agent preferences is conducted with the population size sweeping from 5 to 9 while keeping agents interacting on the exact same map under both Harvest and Cleanup. The individual reward-gathering patterns, as well as the individual aggressiveness patterns, individual contribution effort patterns remain consistent under distinct agent preferences over sub-graphical structures with the increase in population density and lower per-capita resource availability. Under the nearest neighbor preference, the highest mean utilitarian metric is achieved under the Complete topology in Harvest, with a mean of 138.27; the lowest mean utilitarian metric is achieved by the A1 topology, with a mean of 26.98. This illustrates the SRIM model's robustness under increasing population density and non-stationarity. 

Meanwhile, we conducted experiments on the SBM topology with a large population size $N=20$ under both Harvest and Cleanup environment. To avoid resource depletion due to the constraint map, we designed a scaled map with density preservation to ensure per capita resource availability is the same with 5 agent settings for both SSD environments. In Harvest, at agent steps $1e8$, $95\%$ confidence interval of the utilitarian metric of SBM under the nearest neighbor preference is [54.8828, 62.0422], the clique neighbor preference is [47.0825, 48.1625], the critical connection neighbor preference is [64.4392, 72.8421]. The pairwise comparison using the t-test with Bonferroni correction ($\alpha = 0.05$) is conducted for the utilitarian metric under Nearest-Neighbor (NN), Clique-Neighbor (CN) , Critical-Connection neighbor (HBN) for the SBM Topology across 5 seeds. Nearest-Neighbor (NN) vs. Clique-Neighbor (CN) preference with  $t=8.3140$, $p_{\text{adj}}$$=0.0028$,   Nearest-Neighbor (NN) v.s. Critical-Connection neighbor (HBN) preference with  $t=-5.1197$, $p_{\text{adj}}$$=0.0029$, Clique-Neighbor (CN) preference vs. Critical-Connection neighbor (HBN) preference $t=-13.7764$, $p_{\text{adj}}$$=0.0004$. Moreover, The pairwise comparison using the t-test with Bonferroni correction ($\alpha = 0.05$) for BCI metric across different agent preferences are all significantly different. The pairwise comparison using the t-test with Bonferroni correction ($\alpha = 0.05$) is conducted for the BCI metric under Nearest-Neighbor (NN), Clique-Neighbor (CN) , Critical-Connection neighbor (HBN) for the SBM Topology across 5 seeds in Harvest. Nearest-Neighbor (NN) vs. Clique-Neighbor (CN) preference with  $t=-12.3066$, $p_{\text{adj}}$$=0.000005$,   Nearest-Neighbor (NN) v.s. Critical-Connection neighbor (HBN) preference with  $t=-22.2424$, $p_{\text{adj}}$$=0.0000$, Clique-Neighbor (CN) preference vs. Critical-Connection neighbor (HBN) preference $t=-11.9538$, $p_{\text{adj}}$$=0.000014$.A pairwise comparison using the t-test with Bonferroni correction ($\alpha = 0.05$) is conducted for the BCI metric under Nearest-Neighbor (NN), Clique-Neighbor (CN), and Critical-Connection neighbor (HBN) preferences for the SBM Topology across 5 seeds under Cleanup: Nearest-Neighbor (NN) preference vs, Clique-Neighbor (CN) preference with t$=--30.4407$, $p_{\text{adj}}=0.002405$,  Nearest-Neighbor (NN) preference vs. Critical-Connection neighbor (HBN) preference t$= -17.7444$, $p_{\text{adj}}= 0.003509$, Clique-Neighbor (CN) preference vs. Critical-Connection neighbor (HBN) preference t$=-8.2023$, $p_{\text{adj}}=0.042821$. By comparing the Bonferroni-corrected p-value comparison with the significance threshold $\alpha = 0.05$, all comparisons show significance after correction for multiple testing across different agent preferences under SBM with increased population (N=20) under a resource density-controlled environment across both SSDs environments.

The pairwise statistical test on BCI metric and utilitarian metric of the SBM model demonstrates a significant difference across different preferences with increased population under a resource density-controlled environment. The individual reward gathering trajectories are consistent with population size from $N=5$ to $N=9$. Those experimental results illustrate that the model is robust to larger population rates with controlled density and robust to map scaling and modifications. Furthermore, these analyses support our observation that the SRIM model reveals consistent undiscovered strategic patterns over distinct sub-graphical structures under SSDs. Hence, it sets a solid foundation for investigating and understanding agents' strategic behaviors and provides the first controlled, sub-graphical integrated foundation for future study of strategic behaviors and further optimization design on networked SSDs.

\end{document}